\documentclass[aps,prl,reprint,amssymb,showpacs,superscriptaddress,notitlepage,onecolumn]{revtex4-2}
\usepackage{bm}
\usepackage{graphicx,hyperref}
\usepackage{dcolumn}
\usepackage{braket}
\usepackage{natbib}
\usepackage{amsmath}
\usepackage{color}
\usepackage{siunitx}
\usepackage{mathtools,amssymb}
\usepackage{soul}
\usepackage{float}
\usepackage{amssymb}
\usepackage{esint}
\usepackage{mathtools}
\usepackage{physics}
\usepackage{makecell}
\usepackage{multirow}
\usepackage{array}
\usepackage{geometry}
\usepackage{url}
\usepackage{adjustbox}

\begin{document}
\title{Optical pumping of electronic quantum Hall states with vortex light}
\author{Deric Session}
\altaffiliation{These authors contributed equally to this work}
\affiliation{Joint Quantum Institute (JQI), University of Maryland,
College Park, MD 20742, USA}
\author{Mahmoud Jalali Mehrabad*}
\email{mjalalim@umd.edu}
\affiliation{Joint Quantum Institute (JQI), University of Maryland,
College Park, MD 20742, USA}
\author {Nikil Paithankar*}
\affiliation{L-NESS, Department of Physics, Politecnico di Milano, Via Anzani 42, 22100 Como, Italy}
\author{Tobias Grass}
\affiliation{DIPC - Donostia International Physics Center, Paseo Manuel de Lardizábal 4, 20018, San Sebastián, Spain}
\affiliation{Ikerbasque - Basque Foundation for Science, Maria Diaz de Haro 3, 48013, Bilbao, Spain}
\author{Christian J. Eckhardt}
\affiliation{Institut f\"ur Theorie der Statistischen Physik, RWTH Aachen University and JARA-Fundamentals of Future Information Technology, 52056 Aachen, Germany}
\affiliation{Max Planck Institute for the Structure and Dynamics of Matter, Center for Free-Electron Laser Science (CFEL), Luruper Chaussee 149, 22761 Hamburg, Germany}
\author{Bin Cao}
\affiliation{Joint Quantum Institute (JQI), University of Maryland,
College Park, MD 20742, USA}
\author{Daniel Gustavo Su\'arez Forero}
\affiliation{Joint Quantum Institute (JQI), University of Maryland,
College Park, MD 20742, USA}
\author{Kevin Li}
\affiliation{Joint Quantum Institute (JQI), University of Maryland,
College Park, MD 20742, USA}
\author{Mohammad S. Alam}
\affiliation{Joint Quantum Institute (JQI), University of Maryland,
College Park, MD 20742, USA}
\author{Kenji Watanabe}
\affiliation{National Institute for Materials Science,
1-1 Namiki, 305-0044 Tsukuba, Japan}
\author{Takashi Taniguchi}
\affiliation{National Institute for Materials Science,
1-1 Namiki, 305-0044 Tsukuba, Japan}
\author{Glenn S. Solomon}
\affiliation{Department of Physics
University of Adelaide
Adelaide, SA, AU}
\author{Nathan Schine}
\affiliation{Joint Quantum Institute (JQI), University of Maryland,
College Park, MD 20742, USA}
\author{Jay Sau}
\affiliation{Joint Quantum Institute (JQI), University of Maryland,
College Park, MD 20742, USA}
\affiliation{Condensed Matter Theory Center, University of Maryland,
College Park, MD 20742, USA}
\author{Roman Sordan}
\affiliation{L-NESS, Department of Physics, Politecnico di Milano, Via Anzani 42, 22100 Como, Italy}
\author{Mohammad Hafezi}
\email{hafezi@umd.edu}
\affiliation{Joint Quantum Institute (JQI), University of Maryland,
College Park, MD 20742, USA}

%%%%%%%%%%%%%%%%%%%%%%%%%%%%%%%%%%%%%%%%%%%%%%%%%%%%%%%%%%%%%%%%%%%%%%%%%%%%%%%%%%%%%%%%%%%%%%%%%%%%%%%%%%%%%%%%%%%%%%%%%%%%%%

\begin{abstract}
A fundamental requirement for quantum technologies is the ability to coherently control the interaction between electrons and photons. However, in many scenarios involving the interaction between light and matter, the exchange of linear or angular momentum between electrons and photons is not feasible, a condition known as the dipole-approximation limit. An example of a case beyond this limit that has remained experimentally elusive is when the interplay between chiral electrons and vortex light is considered, where the orbital angular momentum of light can be transferred to electrons. Here, we present a novel mechanism for such an orbital angular momentum transfer from optical vortex beams to electronic quantum Hall states. Specifically, we identify a robust contribution to the radial photocurrent, in an annular graphene sample within the quantum Hall regime, that depends on the vorticity of light. This phenomenon can be interpreted as an optical pumping scheme, where the angular momentum of photons is transferred to electrons, generating a radial current, and the current direction is determined by the vorticity of the light. Our findings offer fundamental insights into the optical probing and manipulation of quantum coherence, with wide-ranging implications for advancing quantum coherent optoelectronics.

\end{abstract}

%%%%%%%%%%%%%%%%%%%%%%%%%%%%%%%%%%%%%%%%%%%%%%%%%%%%%%%%%%%%%%%%%%%%%%%%%%%%%%%%%%%%%%%

\maketitle

Coherent manipulation of light-matter hybrids plays a crucial role in advancing future quantum technologies and optoelectronics \cite{bloch2022strongly,basov2016polaritons}. Particularly desirable is the control over the spatial degree of freedom in light-matter interactions. Typically, due to the presence of disorder or Coulomb binding, electronic wavefunctions are much more spatially confined than the wavelengths of associated optical transitions. Consequently, the light-matter interaction occurs locally, and neither the spatial profile of the optical field nor the spatial extent of the electron wavefunction has a significant influence on these interactions, a regime known as the dipole approximation. In other words, in this regime only direct optical transitions are accessible, and the transfer of linear and angular momentum, which enables optical control of the spatial degrees of electrons, is not possible.

To understand this, one can consider a simplified model of a hydrogen-like atom, where the typical Bohr radius ($a_{\text{B}}$) is much smaller than the corresponding optical transition wavelength ($\lambda$). Therefore, the next-order quadrupole transition is weaker by a factor of $(a_{\text{B}}/\lambda)^2$ than the dipole transition, yet it is still observable in experiments \cite{Schmiegelow2016}. One approach to enhance such effects is to shrink the wavelength of the electromagnetic field, which can be achieved by using plasmonic effects \cite{andersen2011strongly,rivera2016shrinking}. Alternatively, if the electronic wave function is coherently extended over the associated optical transition wavelength, and electrons are more itinerant than bound, then a gross violation of the dipole approximation is expected. A striking example is the quantum Hall system, where the electrons in two dimensions are subject to a strong out-of-plane magnetic field. Consequently, the kinetic energy is suppressed and electrons exhibit cyclotron motions with chiral characteristics that make it a promising system to investigate the interplay of the chirality of electrons and photons and the transfer of angular momentum in between \cite{Gullans2017,Takahashi2018,Cao2021,hubener2021engineering,suarez2023spin}.

In particular, there has been a growing interest in investigating chiral and topological effects in photonic systems and also light-matter hybrids \cite{ozawa2019topological,lodahl2017chiral,suarez2023spin,mehrabad2023topological}. Such topological features can be either in the momentum domain and lead to Chern bands, or simply in the spatial degrees of freedom, such as optical vortex beams. Specifically, in addition to spin, in the form of polarization, light can also carry orbital angular momentum (OAM) \cite{allen2003optical, bliokh2023roadmap}. Such an OAM is quantized and given by $\hbar\ell $, where $\hbar$ is the Dirac constant and $\ell$ is the mode number which determines the phase winding of a vortex beam. The interaction of such vortex beams with materials has led to a plethora of exciting phenomena \cite{Quinteiro2022}, such as the orbital photogalvanic effect \cite{Ritesh2020}.

In this work, we experimentally demonstrate the transfer of OAM to electrons in a quantum Hall graphene device with annular geometry using optical vortex beams. In particular, harnessing non-conventional optical selection rules of the Landau levels (LLs) described in Fig.~{\ref{schematic}}c, we show a vorticity-selective light-matter interaction between twisted light and the electronic wavefunctions manifesting as a radial photocurrent (PC). We show that this radial PC only depends on the vorticity of light as a direct indication of spatially coherent light-matter interaction. We provide further evidence of the robustness of this mechanism by comparison with circularly polarized light. Specifically, we find that the PC contribution from OAM is at least one order of magnitude larger than the contribution of spin angular momentum (polarization), allowing us to confirm the significant role of the beam's spatial topology, and its ability to control the spatial degree of electrons.

\section{OAM pumping of electrons}

\begin{figure}
    \centering    
    \includegraphics[width=\columnwidth]{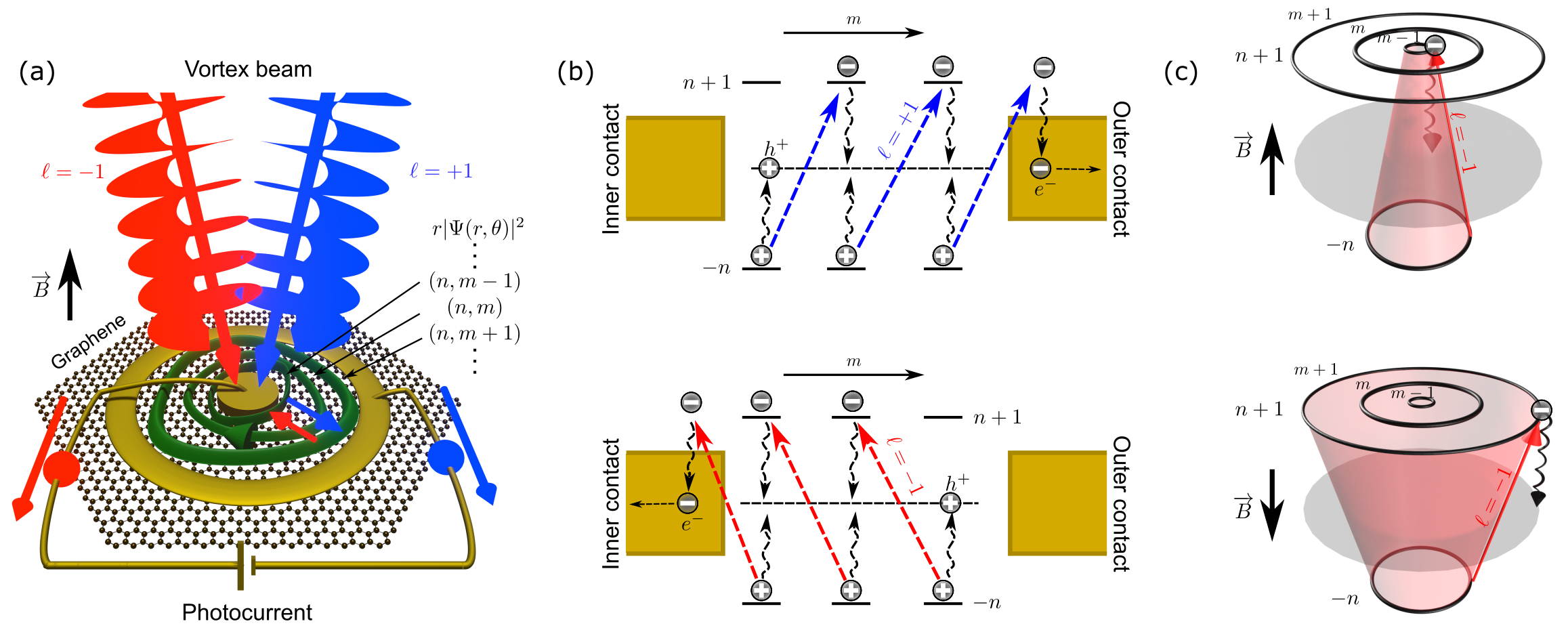}
    \caption{Concept of OAM pumping: (a) Schematic of LLs in the annular disc geometry subject to the vertically irradiated vortex beam. The green surfaces show representative states within a single LL. The electron distributions in the lowest LLs and directions of the induced radial PC are indicated by arrows. (b) Optical transitions and PC generation between inner and outer contacts using $\ell=+1$ (blue arrows) and $\ell=-1$ (red arrows) for $\sigma^+$ polarized light. For $\sigma^-$ transitions see the supplementary information (SI) section S7. Light carrying OAM $\ell=+1$ ($\ell=-1$) increases (decreases) $m$ and hence leads to expansion (shrinking) of the spatial extent of the electronic wavefunction. In this way, the shown scheme realizes an analog to optical pumping. (c) Schematic of the photoexcitation transitions between LLs with negative (red arrow) vortex beam in the presence of an upward (top) and downward (bottom) pointing magnetic field. In the top (bottom) panel, the magnetic field is anti-parallel (parallel) to the helicity of light, leading to the shrinking (expansion) of the wavefunction radius. In (c), only the electron relaxation is considered.
    }
    \label{schematic}
\end{figure}

To present the motivation and a basic understanding of our experiment, we discuss spatially-dependent light-matter interactions that can manipulate the spatial degrees of freedom of electrons within a quantum Hall system. In particular, using a LL picture, we observe how the transfer of OAM from photons to electrons results in a radial current, where the direction of the current is determined by the vorticity of the light. As shown in Fig.~{\ref{schematic}}, we consider optical transitions between two LLs, in the presence of rotational symmetry perpendicular to the plane of the quantum Hall sample. In this scenario, the sample is irradiated by an optical vortex where each photon is carrying an OAM $\hbar \ell$ \mbox{\cite{Gullans2017,Cao2021}}.

During the excitation process, the OAM of $\hbar \ell$ is transferred to electrons \mbox{\cite{Gullans2017,Cao2021}}. As the radius of the electronic wavefunction increases monotonically with angular momentum, this optical transfer of angular momentum causes a radial change in the electronic wavefunction, which is solely determined by the vorticity of the light. The subsequent relaxation process conserves OAM on average and therefore maintains the OAM transfer from the original excitation. This concept is in direct analogy to optical pumping in atomic systems, wherein cyclical pumping among different hyperfine states of bound electrons within an atom transfers them to a specific quantum state \cite{cohen1966optical}. 

Note that despite the absence of rotational symmetry in the presence of disorder, the optical pumping model continues to hold true \cite{Cao2021}. Moreover, the optical pumping picture also provides a simple estimate of the resulting PC: assuming that OAM pumping was the only mechanism of charge transport, the OAM needed to carry one electron through the sample equals the number $M$ of orbitals in a LL, which is given by $M=A/(\pi l_B^2)$, where $A$ is the area of the sample and $l_B$ is the magnetic length. The estimated transported charge, $ \delta Q$, during the time interval $\delta t$ by $N_{\rm ph}$ photons, is $\delta Q = e N_{\rm ph} \ell/M$. Therefore, for photons with energy $\hbar \omega$, the PC obtained from the laser power $P=\hbar \omega \times N_{\rm ph}/\delta t$ is $I = \delta Q/\delta t = e P \ell/(\hbar \omega M)$. This picture also implies that, upon inverting the direction of the magnetic field, the PC changes direction, as illustrated in Fig.~{\ref{schematic}}c.

\begin{figure}
    \centering    
    \includegraphics[width=\columnwidth]{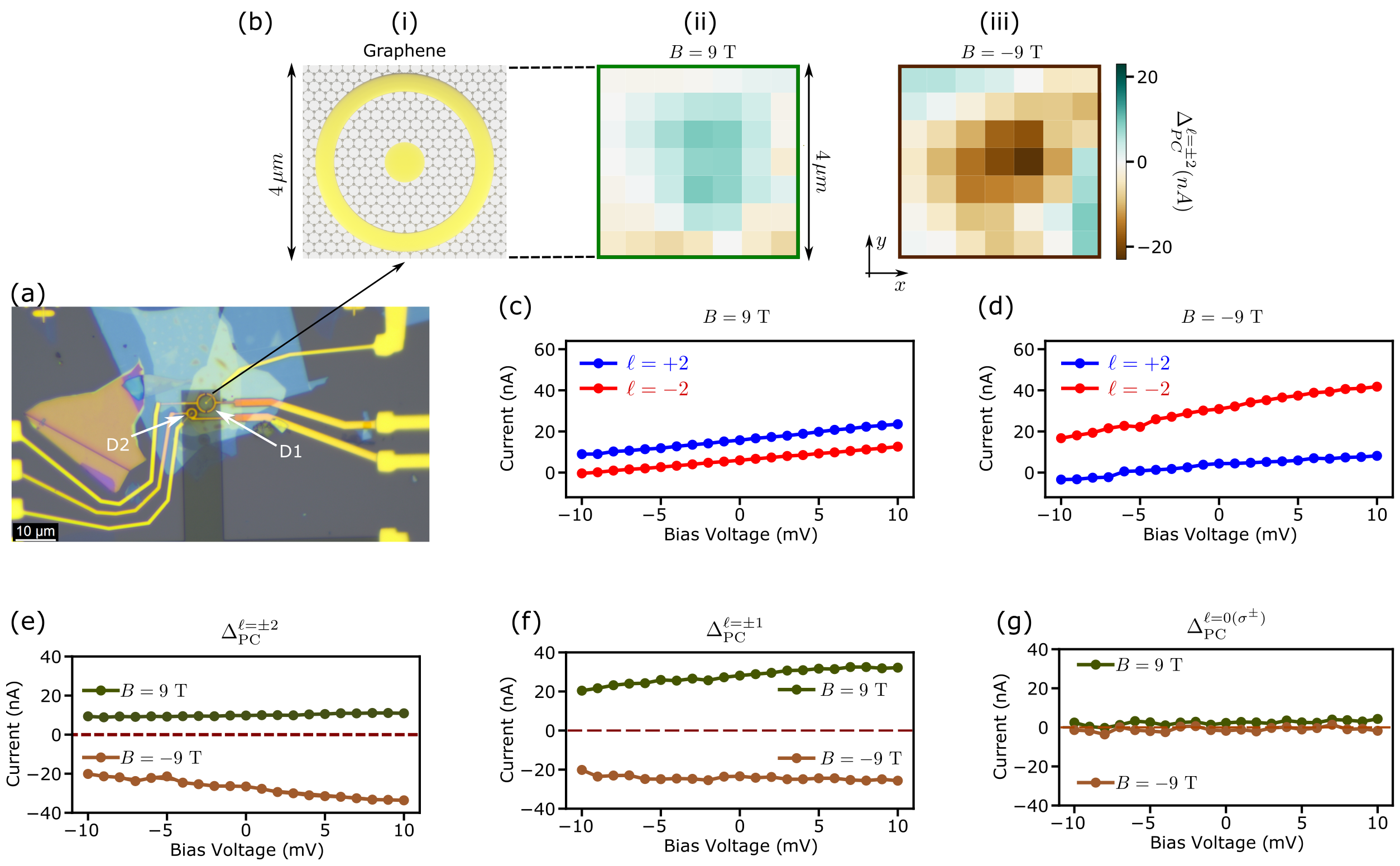}
    \caption{OAM-selective PC generation: (a) Optical-microscope image of the sample, showing two Corbino devices labeled D1 and D2, one with a smaller diameter ($\sim 2\ \rm{\mu m}$) and the other with a larger diameter ($\sim 4\ \rm{\mu m}$). D1 is the device used for the data in the main text of this article, for which the inner and outer contacts are marked. The gray rectangle shows the metallic back gate. (b) (i) Sample schematic, (ii-iii) Spatially-resolved PC difference $\Delta_{\rm{PC}}$ for $\ell=\pm2$, at $B=9\ \rm{T}$ and $B=-9\ \rm{T}$, respectively. Bias voltage $V_{\rm{b}}$ for both cases was $-4\ \rm{mV}$. (c), (d) Measured PC as a function of bias voltage $V_{\rm{b}}$ generated using light carrying $\ell=+2$ (blue) and $\ell=-2$ (red) at (c) $B=9\ \rm{T}$ and (d) $B=-9\ \rm{T}$. (e-g) Shows the $\Delta_{\rm{PC}}$ for $B=9\ \rm{T}$ (green) and $B=-9\ \rm{T}$ (brown) for $\ell=\pm2,\pm1,0$, respectively. In (g), at each magnetic field, the $\Delta_{\rm{PC}}$ is calculated by subtracting the PC generated using two Gaussian beams with opposite circular polarization ($\sigma^\pm$). In all panels, the gate voltage $V_{\rm{g}}$ and the average pump power were $1.78\ \rm{V}$ and $10\ \rm{\mu W}$, respectively. Relative differences of curves within and between panels e and f depend on system parameters, which are described in SI.
    }
    \label{vb}
\end{figure}

\section{Bias voltage dependence of PC}

To experimentally demonstrate this mechanism, we use a device consisting of a hexagonal boron nitride (hBN) encapsulated monolayer graphene in an annular (Corbino) geometry as shown in Fig.~\ref{vb}a. The inner and outer contacts are used to apply an in-plane electric field and also measure the generated PC, while the back-gate voltage $V_{\rm{g}}$ controls the Fermi level. We apply an out-of-plane external magnetic field up to 9 T at 4.2 K to be in the quantum Hall regime. The optical vortex beams of different vortices $\ell$ are generated by a spatial light modulator (SLM) and are concentrically focused on the Corbino device (see the SI for the sample and optical setup, sections S1 and S3). This enables us to excite the carriers in our device which undergo vorticity-selective optical transitions, shown in Fig.~\ref{schematic}c. For all the measurements, we choose $V_{\rm{g}}=\text{1.78 V}$, which sets the Fermi energy near filling factor $\nu=6$  (see Fig.~{\ref{vg}} and associated text for discussion about this choice).

The generated PC for various optical vortices $\pm \ell$ are independently measured, while the beam is spatially scanned over the sample.  Fig.~{\ref{vb}}b shows PC difference (subtracted), $\Delta_{\rm{PC}}$, where the vortex beam $\ell=\pm2$ is spatially scanned over the sample (shown schematically in (i)), at (ii) $B=9\ \rm{T}$ and (iii) $B=-9\ \rm{T}$. It can be clearly observed that the $\Delta_{\rm{PC}}$ flips sign when the magnetic field direction is reversed. This remarkable observation corroborates with the earlier optical pumping picture of  Fig.~\ref{schematic}a, where depending on the vortices of the optical beam the radial extent of the electrons either shrink or expand during the optical excitation. Note that the sign of the observed PC difference depends solely on the phase winding of the optical vortex beam, and not the intensity; therefore, one can not associate this (beyond the dipole-approximation) process with heating.

It is unlikely that any sample has pristine electrical conditions and it may harbor residual or intrinsic in-plane potential. Such inherent potential could potentially explain the presence of a radial PC in the Corbino sample. In order to rule out the origin of our observed effect to such an in-plane electric field, we apply a bias voltage $V_{\rm{b}}$ between the inner and outer contacts to create a controllable potential gradient in the radial direction. Figures~{\ref{vb}}c-d show the measured PC as a function of $V_{\rm{b}}$ for $B=9\ \rm{T}$ and $B=-9\ \rm{T}$, respectively. Remarkably, we observe multiple unambiguous signatures of the vorticity-selective light-matter interaction. First, as shown in Fig.~{\ref{vb}}c-d, for $\ell=+2$ (blue) and $\ell=-2$ (red) at each magnetic field, we observe a consistent and significant difference in the generated PC for a wide range of $V_{\rm{b}}$ ($-10\le V_{\rm{b}}\le10$ mV). In other words, radially tilting the electric potential can change the total PC, however, the PC difference $\Delta_{\rm{PC}}$ remains relatively constant. Second, for the opposite magnetic field, we observe a clear sign flip for the PC difference (Fig.~{\ref{vb}}e). Specifically, since the OAM is defined relative to the magnetic field, inverting the latter effectively inverts the OAM and should therefore lead to a sign-change of the observed radial PC, which is clearly observed in Fig.~{\ref{vb}}e. In an ideal case, the amplitude of this flipped current should be the same, however, due to slightly different spatial alignment for different magnetic fields, the magnitude of the PC is different (See Fig.~{\ref{vb}}b (ii) and (iii))). Third, to investigate the effect of the degree of vorticity on  the generated PC, we illuminate the sample with $\ell=\pm 1$ beams. As shown in Fig.~{\ref{vb}}, robust PC difference $\Delta_{\rm{PC}}^{\ell=\pm 1}$ is observed across a wide range voltage bias, and the sign reversal with the magnetic field is present. For a large sample subject to the optical vortex, one expects that $\Delta_{\rm{PC}}$ to increase with the vorticity degree $\ell$ {\cite{Cao2021}}. However, since our sample and optical vortex spatial profile is comparable, and in particular, the spatial overlap for $\ell=\pm2$ is smaller than that of $\ell=\pm 1$, we observe a reduced $\Delta_{\rm{PC}}$ for the larger value of $|\ell|$. Fourth, in order to rule out the origin of $\Delta_{\rm{PC}}$ based on circular polarization, the sample is illuminated with $\ell=0$ beams consisting of two Gaussian beams with opposite circular polarization ($\sigma^+$ and $\sigma^-$). As shown in Fig.~{\ref{vb}}g, the PC difference for different circular polarization and $\ell=0$ is at least an order of magnitude smaller than the non-zero $\ell$ cases. Therefore, we associate the non-zero radial current with the OAM of light, rather than the spin angular momentum.

%%%%%%%%%%%%%%%%%%%%%%%%%%%%%%%%%%%%%%%%%%%%%%%%%%%%%%%%%%%%%%%%%%%%%%%%%%%%%%%%%%%%%%%%%%%%%%%%%%%%%%%%%%%%%%%%%%%%%%%%%%%%%%
\section{Polarization dependence of PC}

\begin{figure} [t]
    \centering    
    \includegraphics[width=\columnwidth]{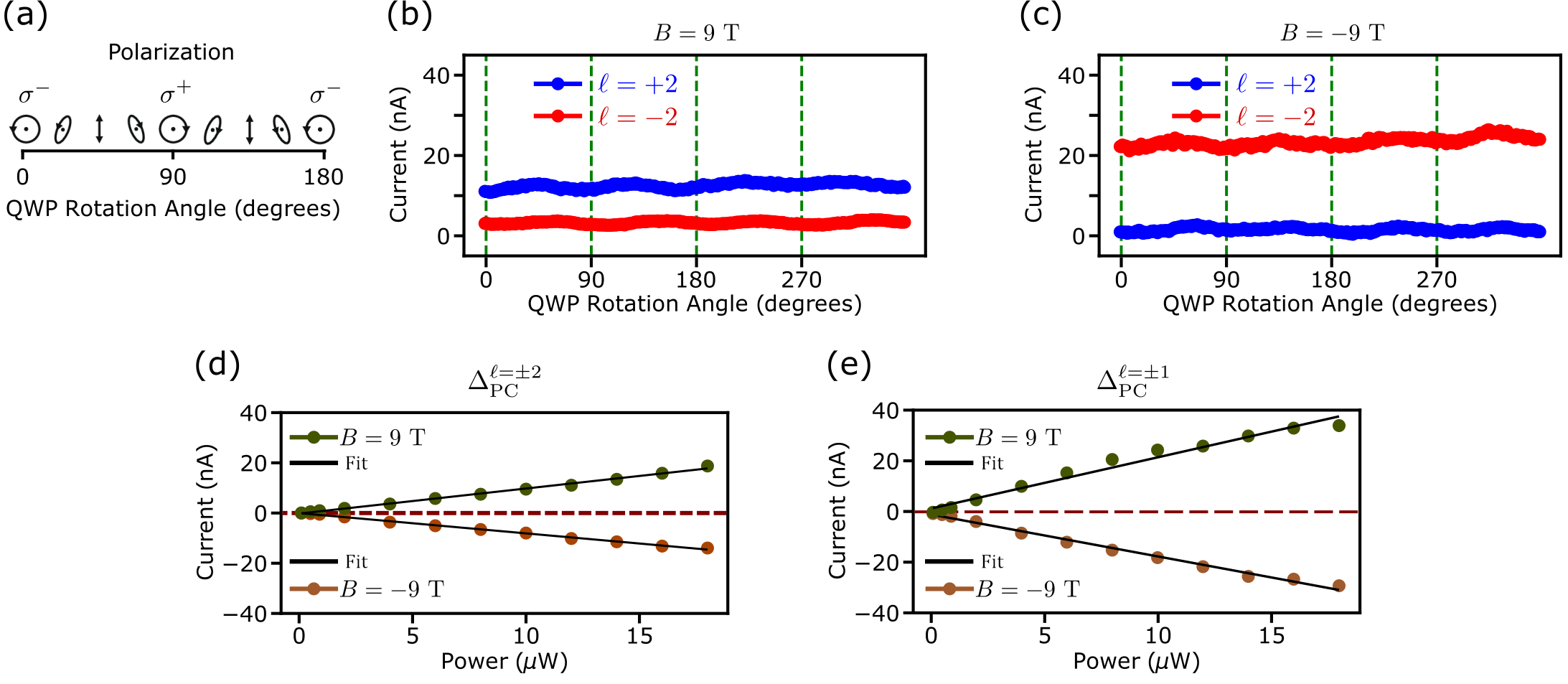}
    \caption{Polarization-resolved and power-dependent PC measurements: (a) polarization of the pump beam as a function of the QWP rotation, shown schematically at the top. $\sigma^+$ and $\sigma^-$ denote positive and negative-circular polarization. Between each consecutive circular polarization, continuous rotation of the QWP changes the polarization periodically from circular to elliptical, to linear polarization. PC measured as a function of QWP rotation using light carrying $\ell=+2$ (blue) and $\ell=-2$ (red) at (b) $B=9\ \rm{T}$ and (c) $B=-9\ \rm{T}$, respectively. The green vertical dashed lines indicate the consecutive $\sigma^+$ and $\sigma^-$ polarization. All measurements are done at a fixed $V_{\rm{g}}=1.78\ \rm{V}$, $V_{\rm{b}}=-4\ \rm{mV}$, and average pump-power of $10\ \rm{\mu W}$. (d), (e) Pump-power-dependence of PC: PC difference $\Delta_{\rm{PC}}$ measured as a function of average pump-power using light carrying (d) $\ell=\pm2$ and (e) $\ell=\pm1$ OAM at $B=9\ \rm{T}$ (green circles) and $B=-9\ \rm{T}$ (brown circles), respectively. The black lines show the least-squares linear fit. In all of the data presented in the main text, the average pump power was set to 10 $\rm{\mu W}$, at which a linear behavior is observed.
    }
    \label{pol}
\end{figure}

To further decouple the role of the polarization of the optical field from OAM in our experiments, we perform polarization-resolved PC generation using OAM of light by using a variable quarter-wave plate (QWP) in the excitation path. We rotate the QWP such that the polarization continuously changes from linear to elliptical to circular while measuring the PC (Fig.~\ref{pol}a). The measured PC for $\ell=+2$ (blue) and $\ell=-2$ (red) at $B=9\ \rm{T}$ and $B=-9\ \rm{T}$ is shown in Fig.~\ref{pol}b-c.

Here, there are three clear observations confirming the robustness of OAM-induced PC generation. First, the OAM-induced PC difference is almost an order of magnitude larger than the amplitude of the current oscillations induced by QWP rotation. Second, the OAM-induced PC difference never changes sign as a function of QWP rotation, further confirming the domination of the observed vorticity-selective PC. Finally, this measurement sheds light on the role of focusing the optical beam on its polarization properties. This is crucial since in our measurements, we focus the beam onto the sample using an aspheric lens with a numerical aperture (NA) of 0.68. This large NA may distort the polarization beyond the paraxial approximation. However, Fig.~\ref{pol}b-c shows that this distortion of polarization is relatively insignificant and it does not affect the validity of our results.

%%%%%%%%%%%%%%%%%%%%%%%%%%%%%%%%%%%%%%%%%%%%%%%%%%%%%%%%%%%%%%%%%%%%%%%%%%%%%%%%%%%%%%%%%%%%%%%%%%%%%%%%%%%%%%%%%%%%%%%%%%%%%%
\section{Power dependence of PC}
In order to verify the linear power dependence of the OAM transfer we also investigate the power dependence of the OAM-induced PC in our measurements as a function of optical pump power by varying the average power of our excitation beam within the range 0.1 to 18 $\rm{\mu W}$. Figures ~\ref{pol}d-e show that the generated PC increases linearly with the pump intensity. From the above optical pumping estimate, we have $l_B\approx 26\ {\rm nm}/\sqrt{B/(1\ \rm tesla)}\approx 9$ nm, and 
$M=A/(\pi l_B^2)\approx 3 \times 10^4$. The PC is estimated to be $I \approx (0.1\ {\rm nA}/{\rm \mu W}) P$. Interestingly, the slope of the experimental power dependence ($\sim 1\ {\rm nA}/{\rm \mu W}$) significantly exceeds this estimate. As we discuss in more detail below, the PC signal also exhibits a strong gate voltage dependence due to relaxation effects, and the experimental values are taken at local PC difference maxima, where significant carrier multiplication can be expected \cite{Cao2022}. In addition, the donut-shaped intensity profile of the light, not taken into account in the estimate, enhances the current flow into the outer contact. Additionally, the OAM-induced PC differences increase linearly with pump power which indicates that our experiment was performed in the linear regime away from the Pauli-blockade in Ref.~\cite{Cao2021}.

%In Ref.~\cite{Cao2021}, the transition between two consecutive Landau levels ($\text{LL}_0$ and $\text{LL}_1$) was considered, while in our experiment, our excitation energy is much higher than the cyclotron energy $\approx 100$ meV and the optical wavelength of 940 nm corresponds to transitions between Landau level of about $n\approx 40$ (see SI for details). Moreover, the weak average excitation pump power used in all of our experiments ($\approx 10\ \rm{\mu W}$) places our result away from any nonlinear regime, indicating that Pauli blocking only has an insignificant effect in our observations.

%%%%%%%%%%%%%%%%%%%%%%%%%%%%%%%%%%%%%%%%%%%%%%%%%%%%%%%%%%%%%%%%%%%%%%%%%%%%%%%%%%%%%%%%%%

\section{Gate voltage dependence of PC}

\begin{figure}
    \centering    
    \includegraphics[width=\columnwidth]{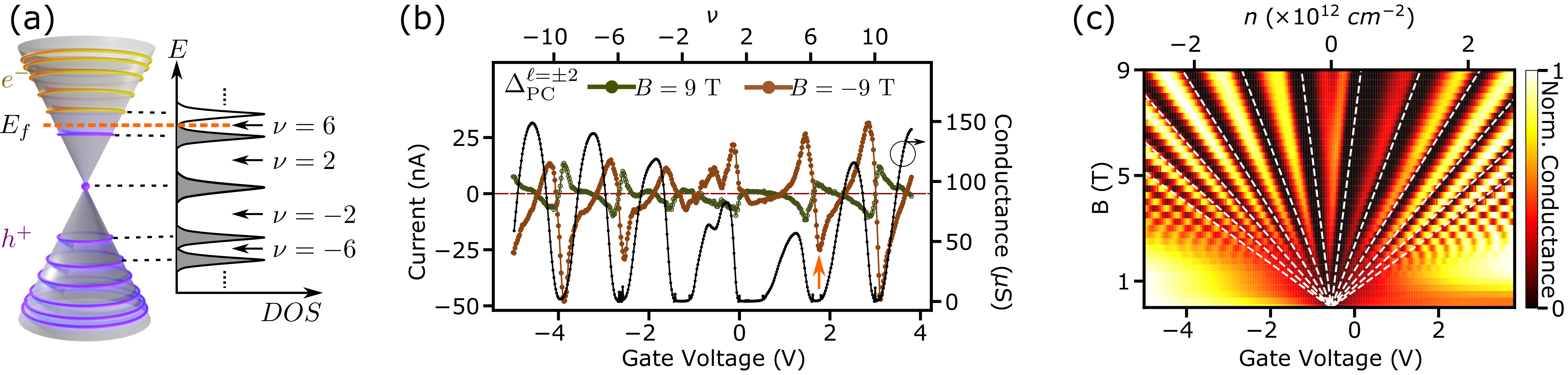}
    \caption{Gate voltage dependence of PC: (a) Schematic of graphene's band structure near the Dirac point with LLs shown in yellow and purple for the electron and holes, respectively. The inset shows the density of states (DOS) as a function of energy with the Fermi energy. (b) Measured PC difference as a function of gate voltage using light carrying $\ell=\pm2$ at $B=9\ \rm{T}$ (green) and $B=-9\ \rm{T}$ (brown), respectively. The top x-axis shows the calculated integer filling factors (see SI section S2). The two-terminal longitudinal conductance measured at $B=9\ \rm{T}$ (black circles) is also shown. This measurement was done at $V_{\rm{b}}=-4\ \rm{mV}$, and an average pump-power of $10\ \rm{\mu W}$. (c) Landau fan measured via magneto-transport as a function of the magnetic field, gate voltage (carrier density marked on the top x-axis). The conductance is normalized for each magnetic field value. The white dashed lines denote integer filling factors calculated from the conversion of gate voltage to carrier density (see SI section S2).
    }
    \label{vg}
\end{figure}

Next, we investigate the role of the Fermi energy ($E_\text{F}$) in the generated vorticity-selective PC. By changing the gate voltage $V_{\rm{g}}$, we tune $E_\text{F}$ between LLs, that is, we tune the LL filling factor $\nu$. As shown in Fig.~{\ref{vg}}b, the subtracted PC (for $\ell=\pm2$) changes sign as a function of $V_{\rm{g}}$, with the direction of PC changing twice between each consecutive conductance peak. In Fig.~\ref{vg}b, green (brown) denotes the subtracted PC at $B=9\ \rm{T}$ ($B=-9\ \rm{T}$), respectively. Recently, an effect known as the ``bottleneck effect", was developed to describe the observation of a similar PC direction change as a function of $V_{\rm{g}}$ in a rectangular geometry {\cite{Cao2022}}. This picture describes this $V_{\rm{g}}$-dependent sign based on the relative position of the electrons and holes compared to the Fermi energy and their propagation through the edge states. However, in contrast to graphene in a rectangular geometry, in our Corbino device, edge states do not contribute to the measured PC. Therefore, our measurements are only sensitive to bulk physics. More discussion about this can be found in the SI.

%%%%%%%%%%%%%%%%%%%%%%%%%%%%%%%%%%%%%%%%%%%%%%%%%%%%%%%%%%%%%%%%%%%%%%%%%%%%%%%%%%%%%%%%%%%%%%%%%%%%%%%%%%%%%%%%%%%%%%%%%%%%%%

\section{Outlook}

In summary, we demonstrated a novel mechanism for transferring OAM from photons to electrons. Our OAM optical pumping scheme is analogous to cold-atom and ion optical pumping, and can be used to manipulate itinerant electrons. Additionally, our results suggest that the control of spatial degrees of freedom in light-matter interactions can become a new and versatile toolbox in solid-state systems. This approach heralds a new ability to image the spatial coherence of electrons, a fundamentally new probe of quantum materials, inaccessible through existing measurements such as multi-port transport and scanning tunneling microscopy \cite{feldman2016observation}. For example, our work highlights the potential of using the OAM degree of freedom as a powerful tool to further the field of quantum Hall physics. One immediate direction is to employ THz fields, as opposed to the optical fields used in our case, to excite the two nearest LLs {\cite{scalari2012ultrastrong}}. The advantage of this approach is the absence of cascade relaxation. Also, the influence of gradient fields on the quantum Hall system has been recently observed in the THz domain, suggesting that this platform is a promising candidate {\cite{appugliese2022breakdown}}. A more ambitious direction involves the strongly interacting limit, where one could exploit the transfer of OAM to probe fractional quantum Hall states and excite and manipulate anyons \mbox{\cite{Tobias2018, knuppel2019nonlinear, ivanov2018adiabatic, binanti2023edge, winter2023fractional}}.

Moreover, our experiment provides a unique testbed for investigating the interplay between topology and chirality in the interactions between electrons and photons. While our experiment was performed in the low excitation limit, there are several intriguing proposals to use a strong drive field and exploit the spatially coherent light-matter interaction to induce a wider class of topological insulators in electronic systems by using various structured light \mbox{\cite{katan2013modulated, Bhattacharya2022, Kim2022, bao2022light, bliokh2023roadmap}}. Furthermore, while our graphene system lacks a photoluminescence response, our scheme can be applied to materials where emission from electronic LLs is possible {\cite{but2019suppressed}}, potentially enabling the observation of chiral photon emission. Another promising avenue is the prospect of coherent wavefunction spectroscopy, where interferometric techniques can be integrated into our experimental scheme to measure and modulate the spatial distribution of wavefunction amplitudes and phases {\cite{zewail2010four}}.

%%%%%%%%%%%%%%%%%%%%%%%%%%%%%%%%%%%%%%%%%%%%%%%%%%%%%%%%%%%%%%%%%%%%%%%%%%%%%%%%%%%%%%%%%%%%%%%%%%%%%%%%%%%%%%%%%%%%%%%%%%%%%%

\section{Acknowledgements}
  The authors acknowledge fruitful discussions with C. Dean, A. Macdonald, I. Kaminer and I. Ahmadabadi. This work was supported by ONR N00014-20-1-2325, AFOSR FA95502010223, ARO W911NF1920181, MURI FA9550-19-1-0399, FA9550-22-1-0339, NSF IMOD DMR-2019444, ARL W911NF1920181, Simons and Minta Martin foundations, and EU Horizon 2020 project Graphene Flagship Core 3 (grant agreement ID 881603). T.G. acknowledges funding by  BBVA Foundation (Beca Leonardo a Investigadores en Física 2023) and  Gipuzkoa Provincial Council (QUAN-000021-01).

\newpage
\section{Supplementary Information: Optical pumping of electronic quantum Hall states with vortex light
}
%%%%%%%%%%%%%%%%%%%%%%%%%%%%%%%%%%%%%%%%%%%%%%%%%%%%%%%%%%%%%%%%%%%%%%%%%%%%%%%%%%%%%%%%%%%%%%%%%%%%%%%%%%%%%%%%%%%%%%%%%%%%%%
\newpage
\section{S1. Device fabrication}

\begin{figure}
    \centering    
    \includegraphics[width=0.99\columnwidth]{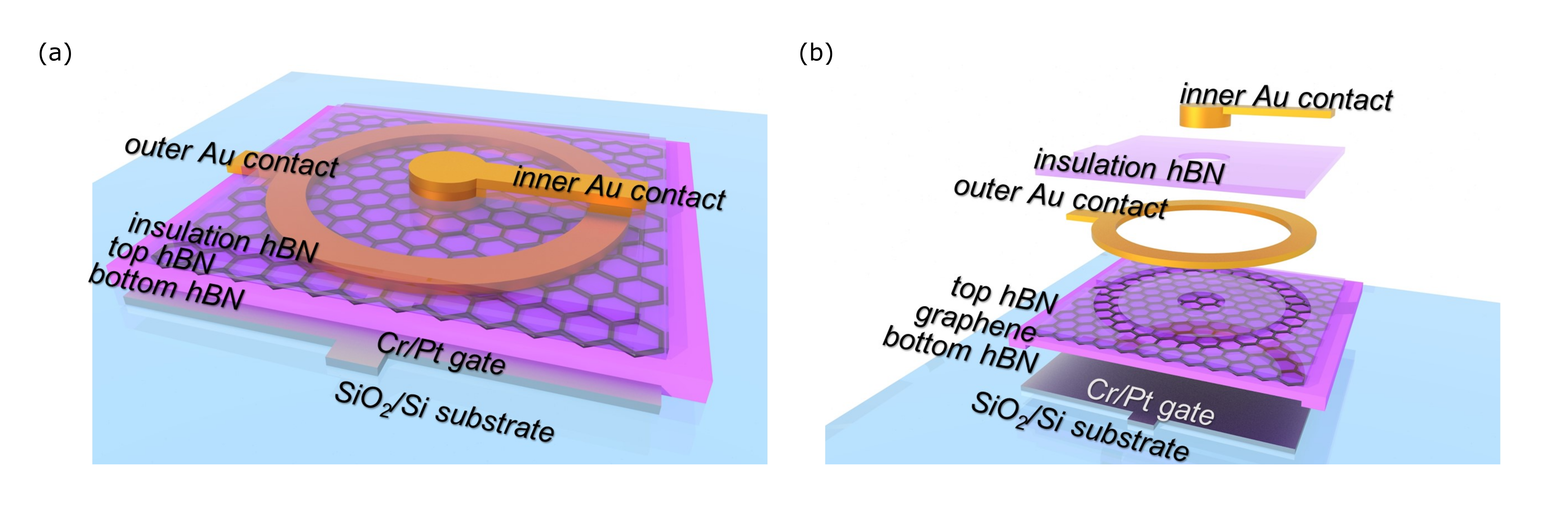}
    \caption{
    Sample fabrication: (a) Schematic of the graphene Corbino sample. (b) Exploded view of (a). 
    }
    \label{fab}
\end{figure}

Graphene was exfoliated from natural graphite crystals (HQ Graphene) and hBN was exfoliated from lab-grown {\cite{Watanabe2004}} or commercial (HQ Graphene) synthetic crystals. Monolayer graphene and hBN were identified based on their color contrast with an optical microscope, while the thickness of the hBN layers used as a gate insulator was measured by atomic force microscopy. A  hot pickup technique was used to stack the hBN/graphene/hBN heterostructure, which was then transferred onto a pre-patterned local metallic back-gate made of 3 nm of Cr and 2 nm of Pt {\cite{Pizzocchero2016}}. The area for the outer contact was etched using a selective reactive ion etching which was followed by the evaporation of 50 nm of Au to form the outer contact {\cite{Jessen2019}}. Next, a third layer of hBN was dropped on top of the heterostructure to act as an insulating layer between the overlapping parts of the contacts. The same selective etch was used to expose the area for the inner contact and 100 nm of Au was evaporated to make the contact. The fabricated devices were wire-bonded to chip carriers for the electrical and photocurrent (PC) measurements.

%%%%%%%%%%%%%%%%%%%%%%%%%%%%%%%%%%%%%%%%%%%%%%%%%%%%%%%%%%%%%%%%%%%%%%%%%%%%%%%%%%%%%%%%%%%%%%%%%%%%%%%%%%%%%%%%%%%%%%%%%%%%%%
\newpage
%%%%%%%%%%%%%%%%%%%%%%%%%%%%%%%%%
\begin{figure}
    \centering    
    \includegraphics[width=0.9\columnwidth]{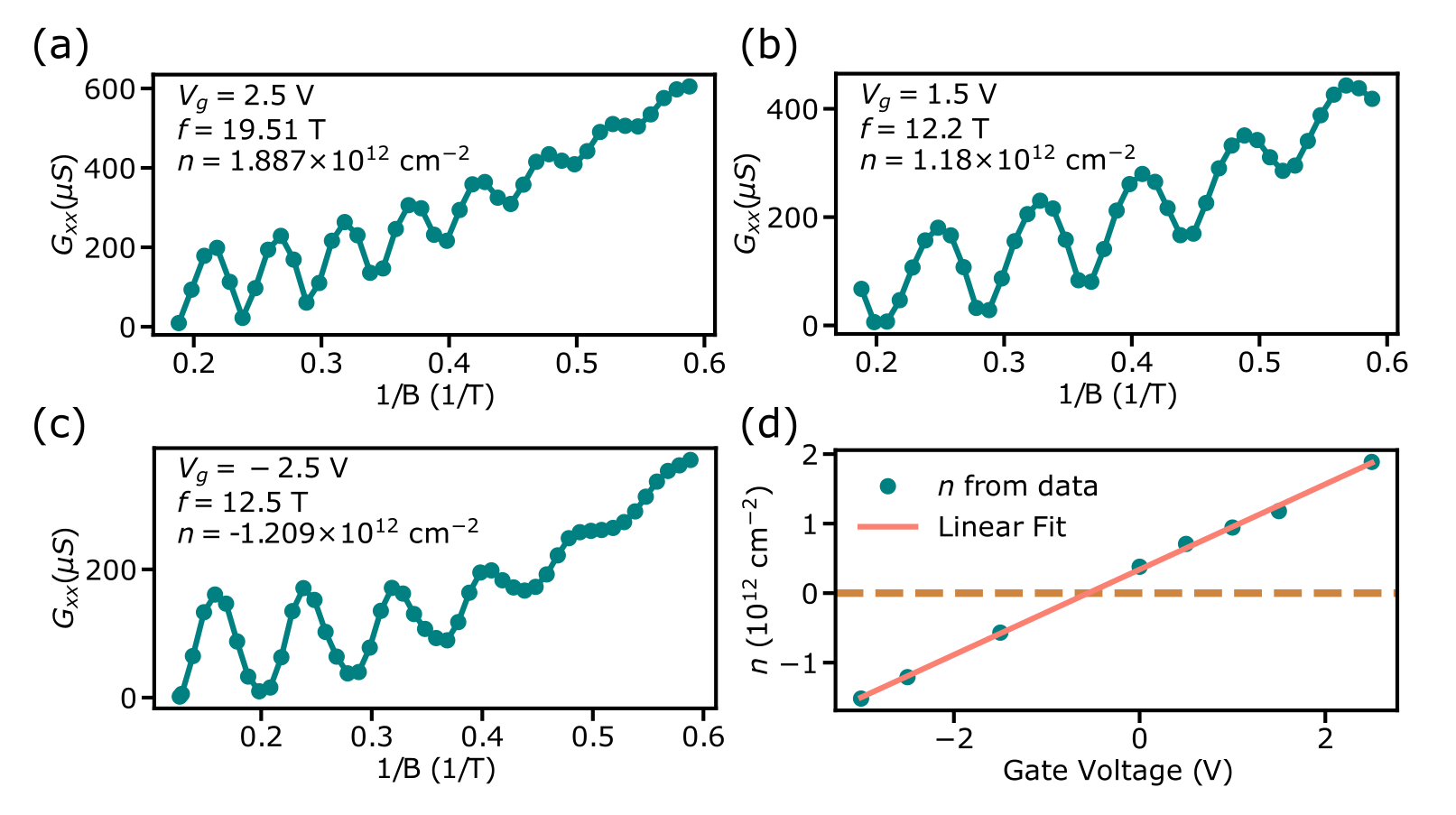}
    \caption{
    Shubnikov-de Haas oscillation measurement: Longitudinal conductance measured as a function of $1/B$ at gate voltages: (a) 2.5 V, (b), 1.5 V, and (c) -2.5 V. Using a Fourier transform to extract the frequency, the carrier densities were calculated to be (a) $1.887\times 10^{12}\ \rm{cm}^{-2}$, (b) $1.18\times 10^{12}\ \rm{cm}^{-2}$, and (c) $-1.209\times 10^{12}\ \rm{cm}^{-2}$. (d) A linear fit to the extracted carrier densities as a function of gate voltage. The slope of the least-squares linear fit is $(0.614 \pm 0.009)\times 10^{12}\ \rm{cm}^{-2}\rm{V}^{-1}$ and the y-intercept is $(0.34 \pm 0.02) \times 10^{12}\ \rm{cm}^{-2}$.}
    \label{SdH}
\end{figure}
%%%%%%%%%%%%%%%%%%%%%%%%%%%%%%%%%
\section{S2. Transport measurements}

\begin{figure}
    \centering    
    \includegraphics[width=0.9\columnwidth]{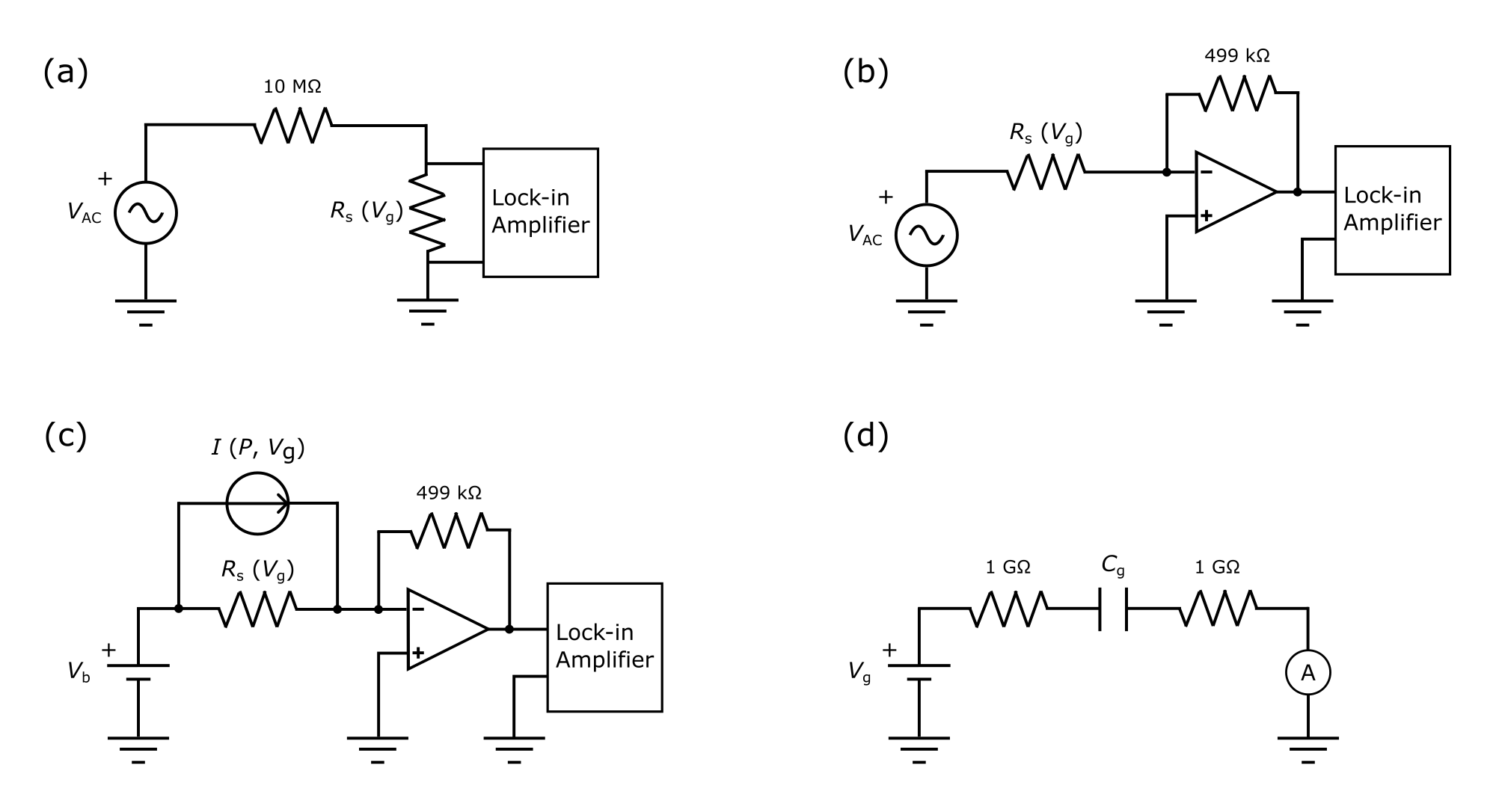}
    \caption{
    Experimental setup: (a) The current bias circuit used for transport measurements. A sinusoidal voltage is provided by the lock-in amplifier. The constant resistor ($10\ \rm{M\Omega}$) is in series with the sample $R_{\rm{S}}$. This large resistor is used to current bias the sample. The voltage across the sample is measured through the inputs of the lock-in amplifier. (b) The voltage bias circuit is alternatively used for transport measurements. We use the same source as in (a), however, there is no large resistor to convert the voltage signal to current. In this case, the resulting current from the sample is passed through a trans-impedance amplifier (TIA), and the resulting voltage is measured with the lock-in amplifier. (c) The effective circuit for measuring PC. In this circuit, the sample is modeled as a current source, $I\ (P,V_{\rm{g}})$, in parallel with a resistor (sample resistance). The sample is in series with a DC voltage source. The TIA is in series with the sample and is connected to both terminals of the lock-in amplifier. (d) The effective circuit used to gate the sample: in this circuit, the sample is treated as a capacitor with capacitance $C_{\rm{g}}$. It is in series with two $1\ \rm{G\Omega}$ resistors to decrease the current in the circuit.}
    \label{circuits}
\end{figure}

Transport measurements are carried out with low-frequency lock-in techniques either by biasing with a current of 20 nA or with a voltage at a frequency of 13 Hz. Gate voltage sweeping is performed using a DC source measure unit (Keithley 2450). The two-terminal longitudinal conductance for various magnetic fields was measured to obtain the Landau fan (Fig. 4c in the main text). These measurements are performed using the circuits depicted in Fig.~\ref{circuits}a-b.

To calibrate the conversion between gate voltage and carrier density, we measured Shubnikov-de Haas oscillations by setting the gate voltage and sweeping the magnetic field. The carrier density is extracted from the frequency of the oscillation ($f$) by $n=4ef/h$ \cite{Zeng2019}. We extracted the carrier density for gate voltages between -3 V and 2.5 V and used a linear fit as a conversion between gate voltage and carrier density, see Fig.~\ref{SdH}.

%%%%%%%%%%%%%%%%%%%%%%%%%%%%%%%%%

\begin{figure}
    \centering    
    \includegraphics[width=\columnwidth]{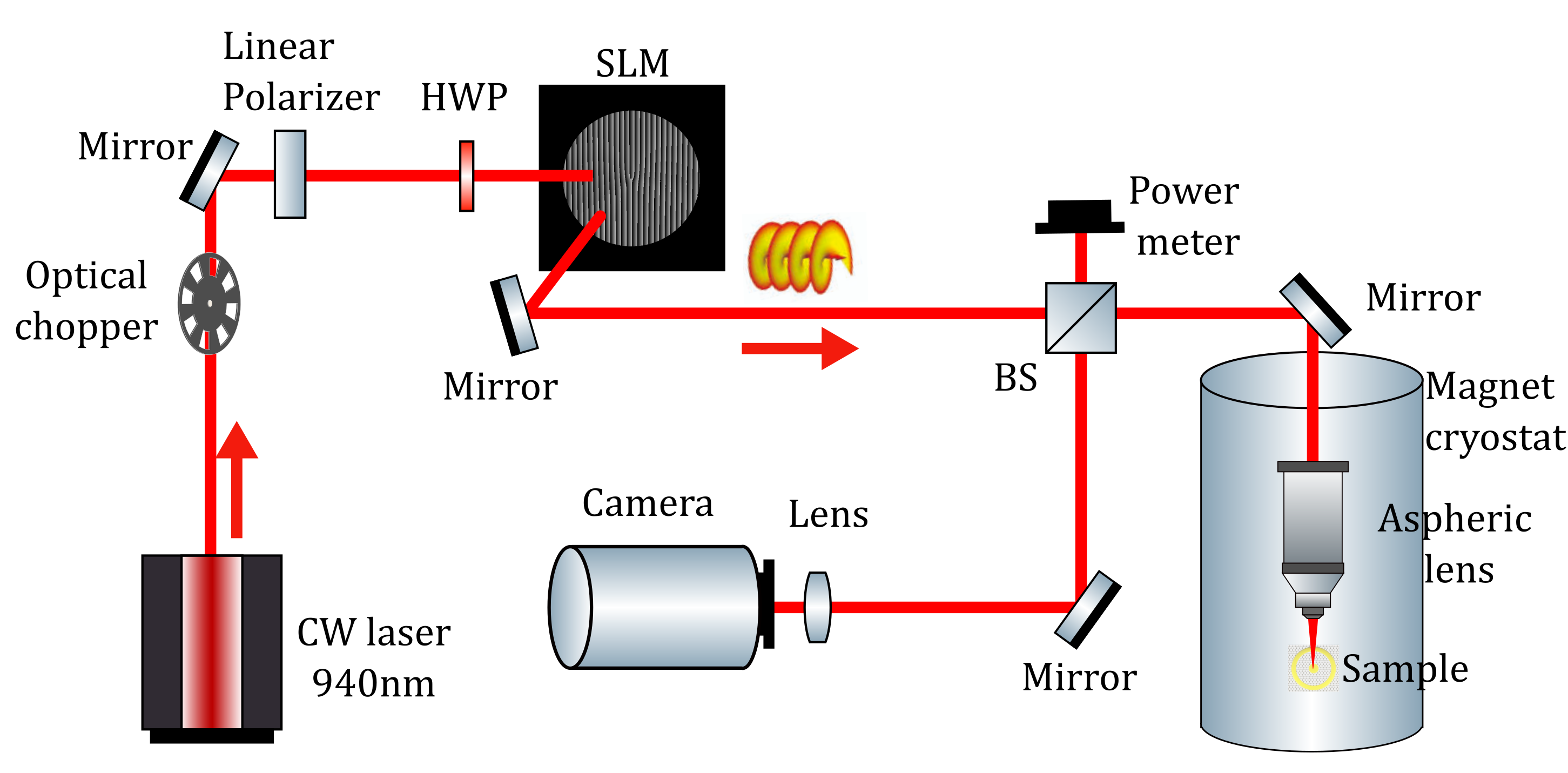}
    \caption{
    The optical setup used for the PC generation.
    }
    \label{Optics}
\end{figure}

%%%%%%%%%%%%%%%%%%%%%%%%%%%%%%%%%%%%%%%%%%%%%%%%%%%%%%%%%%%%%%%%%%%%%%%%%%%%%%%%%%%%%%%%%%%%%%%%%%%%%%%%%%%%%%%%%%%%%%%%%%%%%%
\newpage
\section{S3. PC measurements}

The sample, inside a variable temperature insert (VTI), is mounted on top of a piezo-electric stack (scanners (ANSxy100) and positioners (ANPx101, ANPz201)). It is cooled down to 4.2 K using liquid helium and can reach magnetic fields up to 9 T. The VTI has an optical window on top and a confocal microscope is built above to optically resolve the sample. The pump laser is illuminated through the same window and the laser spot's alignment to the sample can be monitored with the microscope. During measurements, the laser power is constantly monitored with a power meter right before the beam enters the optical window, and a feedback is given to a proportional-integral-derivative loop controlling a laser power control module. The power control module consists of a DC voltage applied to a Thorlabs electronic variable optical attenuator.

The two sample contacts are connected to a custom-built trans-impedance amplifier (TIA) outside of the cryostat \cite{Cao2022}. The outputs of the TIA are connected to a lock-in amplifier (SRS SR860). The pump laser is chopped at a frequency of 308 Hz and the lock-in is frequency-locked to the chopper.

We sweep the gate in the same way as in the transport measurement. However, in the PC measurements, we apply a DC voltage bias across the sample using an SRS SIM928. The PC measurements are performed using the circuit depicted in Fig.~\ref{circuits}c and the optical setup depicted in Fig.~\ref{Optics}.

%%%%%%%%%%%%%%%%%%%%%%%%%%%%%%%%%%%%%%%%%%%%%%%%%%%%%%%%%%%%%%%%%%%%%%%%%%%%%%%%%%%%%%%%%%%%%%%%%%%%%%%%%%%%%%%%%%%%%%%%%%%%%%
\newpage
\begin{figure}[t]
    \centering    
    \includegraphics[width=\columnwidth]{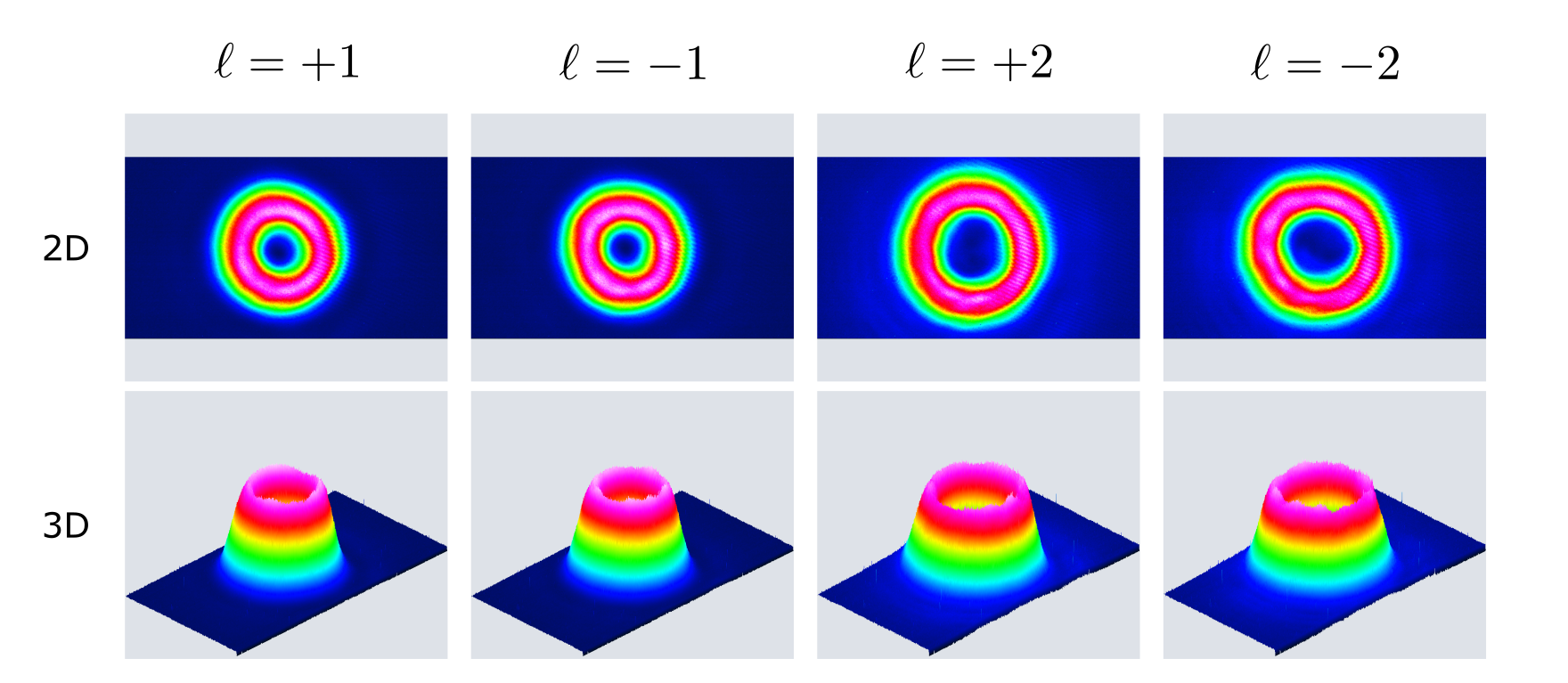}
    \caption{
    Beam profiles are imaged with a beam profiler for $\ell=+1$, $\ell=-1$, $\ell=+2$, and $\ell=-2$. Images are taken of the collimated beam diverted from the sample path.
    }
    \label{BeamProfiles}
\end{figure}

\section{S4. Generation and optimization of orbital angular momentum beams}

To generate beams with orbital angular momentum (OAM) a Gaussian beam is diffracted off of a phase-only spatial light modulator (SLM). Ideally, the pattern displayed to achieve this would be
\begin{equation}
    h(r,\theta)=\frac{1}{2\pi}\rm{mod}\left(\ell \theta - \frac{2\pi}{D} r \cos(\theta),\ 2\pi \right).
\end{equation}
This blazed hologram maximizes the diffraction efficiency \cite{clifford1998high,leach2010quantum}. The obtained diffraction modes are a superposition of Laguerre-Gaussian (LG) modes with the same $\ell$ but various $p$ which is the transverse index of the LG mode \cite{arlt2000applications}. A Meadowlark optics 1920x1152 XY Phase series SLM is used. The performance of the pixels in the SLM is not homogeneous. The phase hologram can be modified to compensate for this. The phase hologram used in the experiment is
\begin{equation}
    h(r,\theta) = \mathcal{M}(r,\theta) \rm{mod} \left(\mathcal{F}(r,\theta) + \frac{2\pi}{D} r \cos(\theta),\ 2\pi \right),
\end{equation}
where $\mathcal{M} = 1 + \frac{1}{\pi}\rm{sinc}^{-1}(A(r,\theta))$ and $\mathcal{F} = \Phi - \pi\mathcal{M}$. The beam can be optimized by setting $\Phi=\ell\theta$ and manually adjusting the function $A(r,\theta)$ until a satisfactory beam is reached \cite{davis1999encoding,bolduc2013exact}. The OAM beams after optimization are shown in Fig.~\ref{BeamProfiles}.

\begin{figure}
    \centering    
    \includegraphics[width=0.6\columnwidth]{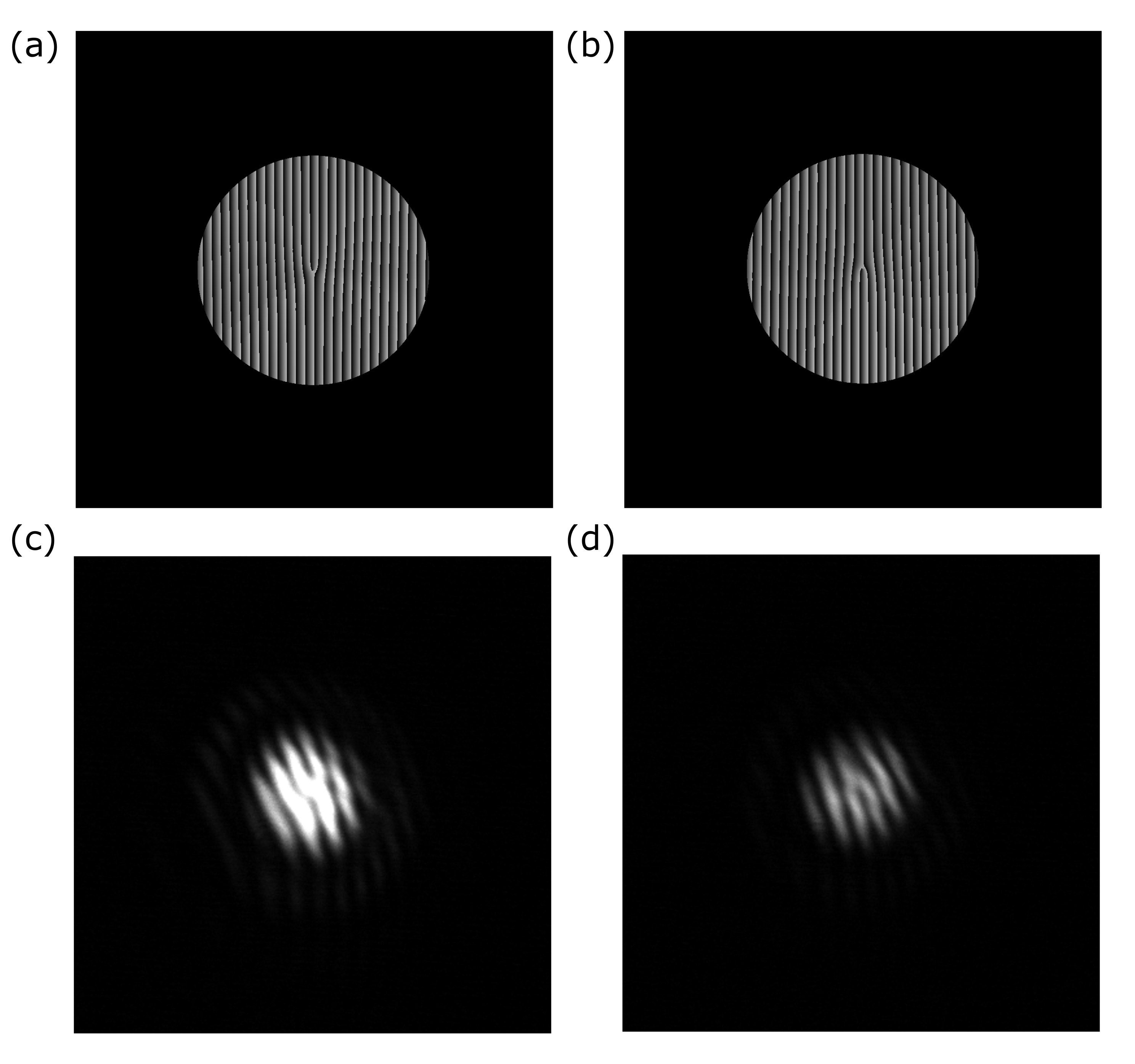}
    \caption{
    (a) Grating displayed on SLM to generate $\ell=+1$ as the first diffraction order. (b) The same as (a), but for $\ell=-1$. (c) Interference pattern when the $\ell=+1$ beam interferes with a Gaussian beam on the sample substrate. (d) The same as (c), but with $\ell=-1$.
    }
    \label{Interference}
\end{figure}

To verify that the beams used in the experiment have the correct OAM, the beam's reflection off of the sample was interfered with a Gaussian reference beam. The results can be seen in Fig.~\ref{Interference}. The interference patterns show the same pitchfork but in opposite directions, verifying that one has $\ell=+1$ and the other has $\ell=-1$.

\subsection{S4a. Profile Intensity Comparison}

To minimize the beam profile intensity mismatch effects in the PC measurements, beam optimization was performed to make the beams with opposite OAMs as similar as possible. However, naturally, the beams will not be perfectly identical, as can be seen in Fig.~\ref{BeamProfiles}. This is important since differing beam profiles can cause a difference in generated PC. To quantify and demonstrate the insignificance of this effect in our experiments, the overlap of the beams is calculated using the following function:
\begin{equation}
    O[I(x,y),J(x,y)] = \frac{\left|\int\int I(x,y)J(x,y)dxdy \right|^2}{\int\int \left|I(x,y)\right|^2dxdy \int\int \left|J(x,y)\right|^2dxdy},
\end{equation}
where $I(x,y)$ and $J(x,y)$ are the two beam profiles with opposite OAMs. The maximum possible overlap is one. The calculated overlap for the $\ell=\pm1$ and $\ell=\pm2$ beams is 0.96 and 0.92 respectively. In other words, the $\ell=\pm1$ beams have a 4\% difference and the $\ell=\pm2$ beams have an 8\% intensity profile difference. The minimum percent difference between the average total current and the subtracted current for $\ell=\pm2$ at $B=9\ \rm{T}$ is 60\%. For $B=-9\ \rm{T}$ it is 134\%. For $\ell=\pm1$ the minimum percent differences are 78\% and 75\% respectively. This confirms that the beam profiles' intensity mismatch cannot be affecting our observations.

Moreover, the beam profiles do not change with flipping the magnetic field, further confirming that the intensity profile mismatch cannot cause the sign of the PC difference to flip for the opposite magnetic field.

%%%%%%%%%%%%%%%%%%%%%%%%%%%%%%%%%%%%%%%%%%%%%%%%%%%%%%%%%%%%%%%%%%%%%%%%%%%%%%%%%%%%%%%%%%%%%%%%%%%%%%%%%%%%%%%%%%%%%%%%%%%%%%

\clearpage

%%%%%%%%%%%%%%%%%%%%%%%%%%%%%%%%%%%%%%%%%%%%%%%%%%%%%%%%%%%%%%%%%%%%%%%%%%%%%%%%%%%%%%%%%%%%%%%%%%%%%%%%%%%%%%%%%%%%%%%%%%%%%%

\section{S5. Gate voltage sweep measurements for different bias}

The gate voltage dependence for different bias voltages was also measured. It is observed that the gate voltage dependence changes very little with bias voltage and the flip with magnetic field stays the same.

\begin{figure}[h]
    \centering    
    \includegraphics[width=\columnwidth]{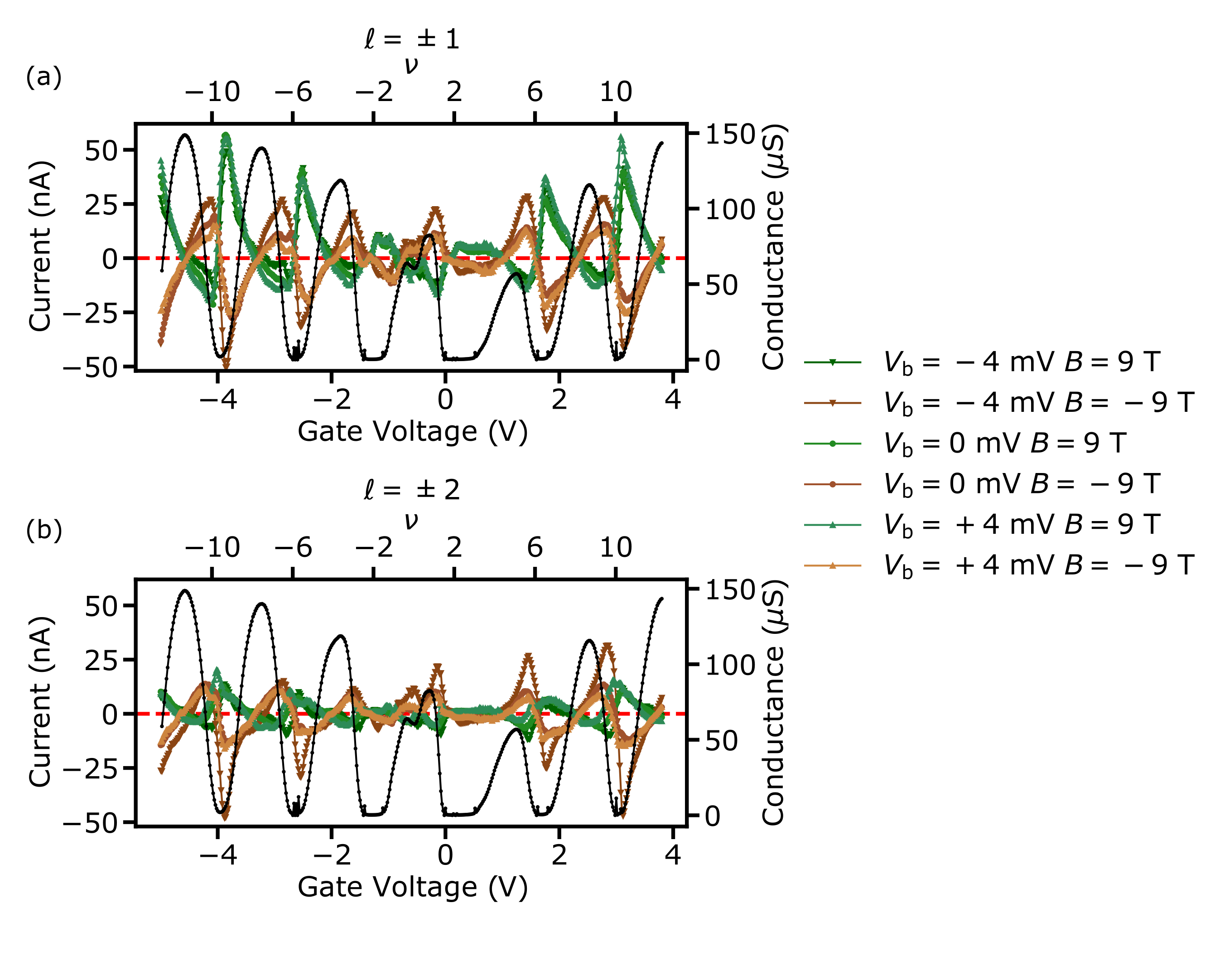}
    \caption{
    Measured PC difference as a function of gate voltage for different $V_{\rm{b}}$ at $B=9\ \rm{T}$: (a) Subtraction of PC for $\ell=\pm1$. (b) Subtraction of PC for $\ell=\pm2$.
    }
    \label{Vg_sweep}
\end{figure}

%%%%%%%%%%%%%%%%%%%%%%%%%%%%%%%%%%%%%%%%%%%%%%%%%%%%%%%%%%%%%%%%%%%%%%%%%%%%%%%%%%%%%%%%%%%%%%%%%%%%%%%%%%%%%%%%%%%%%%%%%%%%%%
\newpage

\section{S6. Reflection Imaging of the Sample}

\begin{figure}[h]
    \centering    
    \includegraphics[width=\columnwidth]{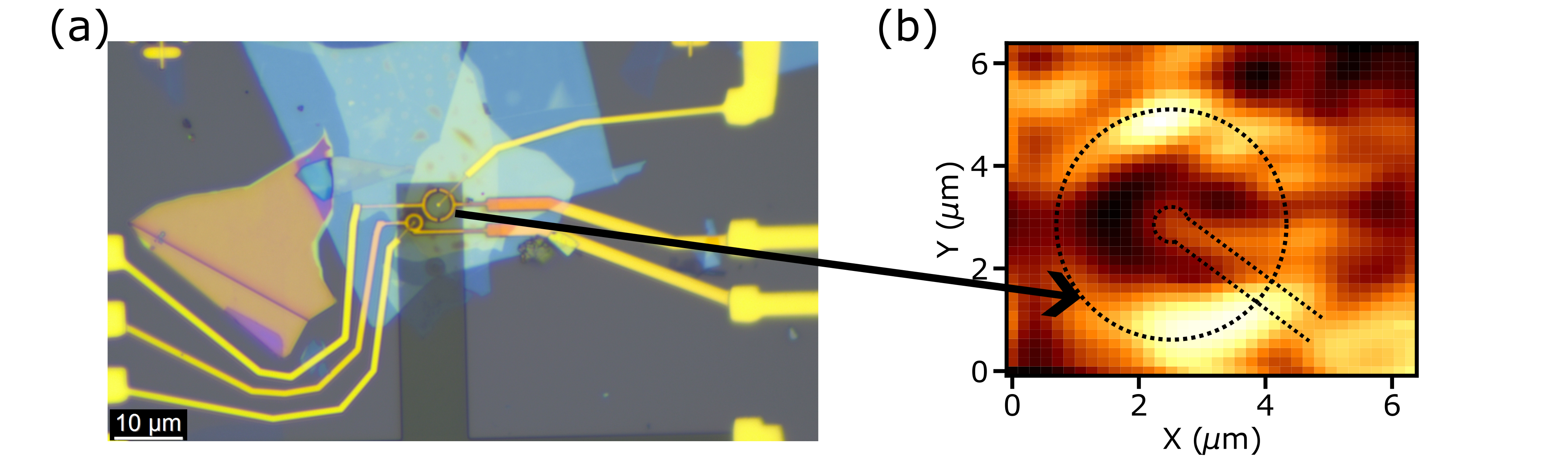}
    \caption{
     (a) Image of the sample taken with an optical microscope. (b) Image of the sample taken by measuring the optical reflection of the pump beam with a photodiode while moving the sample with scanners. The dashed curves show the outline of the Corbino sample.
    }
    \label{reflection}
\end{figure}

In the PC measurement setup depicted in Fig.~{\ref{Optics}}, it is difficult to obtain a high enough quality image of the sample to properly align the beam to it. Another method is to use a Gaussian beam, the same used for PC measurements with $\ell=0$, and move the sample using scanners (ANSxy100) while collecting the reflection with a photodiode. The measured voltage from the photodiode can be used to make an image of the sample; one of these images is shown in Fig.~{\ref{reflection}}b. The gold contacts on the sample are highly reflective in comparison to the hBN/graphene/hBN heterostructure, therefore, in the image, the gold contacts will appear bright and areas of the sample without gold will appear dark (with the chosen color map). Since the outer boundary of the Corbino sample is covered with a gold contact and the center of the Corbino also has a gold contact, the image clearly defines the boundaries of the sample which can then be used for beam alignment.

%%%%%%%%%%%%%%%%%%%%%%%%%%%%%%%%%%%%%%%%%%%%%%%%%%%%%%%%%%%%%%%%%%%%%%%%%%%%%%%%%%%%%%%%%%%%%%%%%%%%%%%%%%%%%%%%%%%%%%%%%%%%%%
\newpage

\section{S7. Selection Rules}
 We discuss the selection rules of exciting electrons with light from the lower to the upper band in graphene in the Quantum Hall regime, considering the possibility that the light might carry non-zero OAM.  
We consider the electromagnetic vector potential $\vec{A}$ of incident light coupled via Peierls substitution $\vec{k} \rightarrow \vec{k} - \mathrm{e} \vec{A}$
and compute transition matrix elements as
\begin{equation}
    M(n_{\mathrm{f}}, m_{\mathrm{f}}, n_\mathrm{i}, m_\mathrm{i}) = - \mathrm{e} v_{\rm F} \langle n_{\rm f}, m_{\rm f} | \vec{A}(\vec{r}) \cdot \vec{\sigma} |n_{\rm i}, m_{\rm i} \rangle
\end{equation}
where $|n_{\rm x}, m_{\rm x} \rangle$ denotes initial (x = i) and final (x = f) states. $n_{\rm x}$ denotes the LL and $m_{\rm x}$ the quantum number related to angular momentum.
The expected angular momentum in $z$-direction can be computed as $\langle L_z \rangle = \hbar( m - n)$.
We compute the matrix elements in the spinor representation of the wave-function limiting to transitions from the lower band of graphene (negative $n$) to the upper band (positive $n$) and assuming $n \neq 0$.
For $\sigma^+$ polarized light, $\vec{A}(\vec{r}) \cdot \vec{\sigma} = A(\vec{r}) \sigma^+$, we get
\begin{equation}
    \begin{aligned}
    M^+(n_{\mathrm{f}}, m_{\mathrm{f}}, n_\mathrm{i}, m_\mathrm{i}) &= - \frac{\mathrm{e} v_{\rm F}}{2} \begin{pmatrix} \langle n_{\rm f} - 1, m_{\rm f}|, & \langle n_{\rm f}, m_{\rm f}  | \end{pmatrix} A(\vec{r}) \sigma^+ \begin{pmatrix} |n_{\rm i} - 1, m_{\rm i} \rangle \\ - |n_{\rm i} , m_{\rm i} \rangle \end{pmatrix}\\
    &= \frac{\mathrm{e} v_{\rm F}}{2} \langle n_{\rm f} - 1, m_{\rm f}| A(\vec{r})  |n_{\rm i} , m_{\rm i} \rangle
    \end{aligned}
\end{equation}
while for $\sigma^-$ polarized light we get
\begin{equation}
     M^-(n_{\mathrm{f}}, m_{\mathrm{f}}, n_\mathrm{i}, m_\mathrm{i}) = - \frac{\mathrm{e} v_{\rm F}}{2} \langle n_{\rm f}, m_{\rm f}| A(\vec{r})  |n_{\rm i} -1 , m_{\rm i} \rangle.
\end{equation}
%In the case of a spatially uniform vector potential $\vec{A}(\vec{r}) = A$ we get the well known selcetion rules for $\sigma^+$ cite:
%\begin{equation}
%    M^+ \sim \delta_{n_{\rm f} - 1, n_{\rm i}} \delta_{m_{\rm f}, m_{\rm i}}
%\end{equation}
%and for $\sigma^-$
%\begin{equation}
%    M^- \sim \delta_{n_{\rm f}, n_{\rm i} - 1} \delta_{m_{\rm f}, m_{\rm i}}.
%\end{equation}

\begin{figure}[h]
    \centering 
\includegraphics[width=\columnwidth]{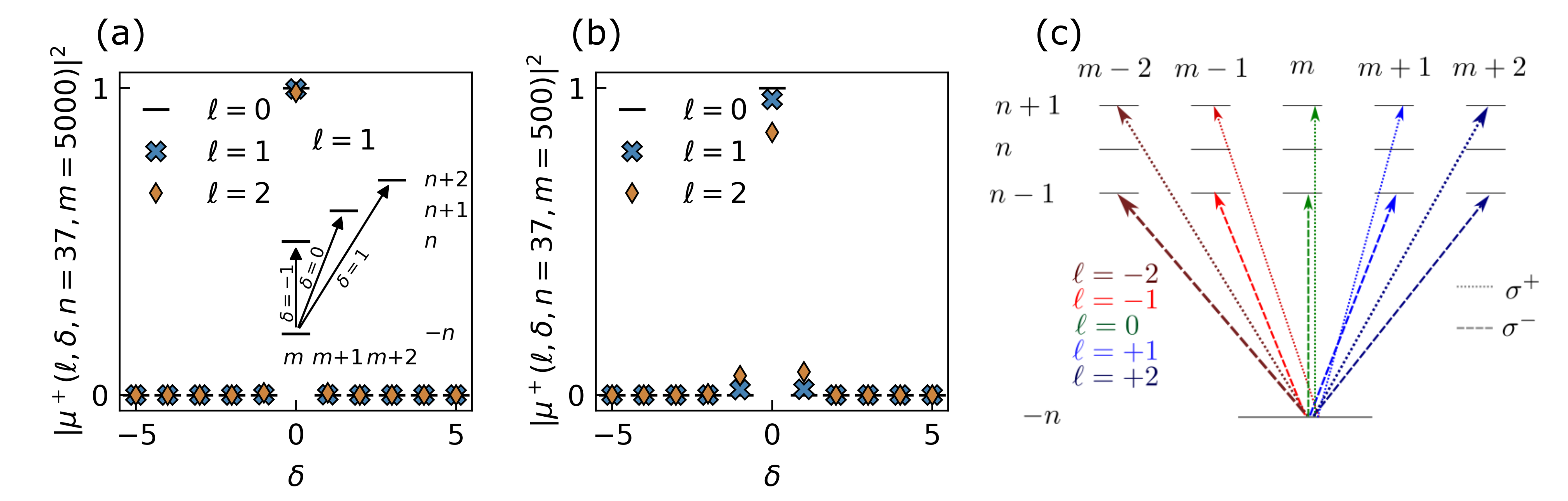}
    \caption{
    (a) Matrix element $\mu^+(\ell, \delta, n, m)$ as defined in Eq.~(\ref{eq:marixElementWrittenOut}) as a function of $\delta$ for $\ell = 0, 1, 2$ as in the experiment (limiting to positive OAM). $n = 37$ is fixed as in the experiment and $m=5000$ which corresponds to a state with expected radial position $\langle r \rangle_{n = 37, m=5000} \approx 1\ \mu \mathrm{m}$ which is somewhere in the middle of the sample, with radius $R = 2\ \mu \mathrm{m}$. The inset illustrates the effect of $\delta$ in the case of $\ell = 1$: $\delta = 0$ corresponds to the situation where the OAM of the light only changes the quantum-number $m$ while $\delta \neq 0$ considers further transitions where $n$ and $m$ are simultaneously changed. Only the $\delta = 0$ transition has substantial weight.
    (b) Same as in $a.)$ just with $m = 500$ which corresponds to a radial position of the electron of $\langle r \rangle_{n = 37, m=500} \approx 300\ \mathrm{nm}$ hence close to the inner contact. The $\delta = 0$ transitions also dominate here but for smaller $m$, $\delta \neq 0$ transitions have an increased weight.
    \textbf{c.)} Illustration of selection rules considering only $\delta = 0$ that have the dominant weight (see (a) and (b)).
    }
    \label{fig:SelectionRules}
\end{figure}

For light carrying OAM, we write $A(\vec{r}) = A e^{i \varphi \ell}$, $\ell = 0, \pm 1, \pm 2, \dots$ assuming no radial dependence of the light field.
We compute the dimensionless and normalized matrix elements for $\sigma^+$ polarized light
\begin{equation}
    \mu^+ = \langle n_{\rm f} - 1, m_{\rm f} | e^{i \varphi \ell} | n_{\rm i}, m_{\rm i} \rangle
    \label{eq:muPlus}
\end{equation}
and $\sigma^-$ polarized light
\begin{equation}
    \mu^- = \langle n_{\rm f}, m_{\rm f} | e^{i \varphi \ell} | n_{\rm i} - 1, m_{\rm i} \rangle
    \label{eq:muMinus}
\end{equation}
using the wave-functions
\begin{equation}
    \langle r, \varphi | n, m \rangle = \frac{i^{(m - n)}}{\sqrt{2 \pi} \ell_{\rm B}} \sqrt{\frac{n!}{m!}} e^{-\frac{r^2}{4 \ell_{\rm B}^2}} e^{i \varphi (m - n)} r^{(m - n)}  L_{n}^{(m - n)} \left(\frac{r^2}{2 \ell_{\rm B}^2}\right).
    \label{eq:waveFunction}
\end{equation}
Here $L_{a}^{b}(x)$ are the associated Laguerre polynomials.
We assume $n < m - 1$, since this is the relevant case for the experiment.
In what follows, we will focus on $\sigma^+$ polarized light computing $\mu^+$.
However, analogous steps can be performed to compute the matrix elements $\mu^-$.
Inserting the wave function \cite{goerbig_electronic_2011, wendler_ultrafast_2015} Eq.~(\ref{eq:waveFunction}) into Eq.~(\ref{eq:muPlus}) we get
\begin{equation}
\begin{aligned}
    \mu^+ =& i^{(m_{\rm i} - m_{\rm f} - n_{\rm i} + n_{\rm f} - 1)}  \sqrt{\frac{n_{\rm i}! (n_{\rm f} - 1)!}{m_{\rm i}! m_{\rm f}!}} \frac{1}{2 \pi} \int_0^{2\pi} \mathrm{d}\varphi \, e^{i \varphi (m_{\rm i} - m_{\rm f} - n_{\rm i} + n_{\rm f} - 1 + \ell)} \\
    &\int_0^{\infty} \mathrm{d}x \, 2 x e^{-x^2} x^{m_{\rm i} + m_{\rm f} - n_{\rm i} - n_{\rm f} + 1} L_{n_{\rm i}}^{m_{\rm_i} - n_{\rm i}}(x^2) L_{n_{\rm f} - 1}^{m_{\rm_f} - n_{\rm f} + 1}(x^2).
\end{aligned}
\label{eq:marixElementWrittenOut}
\end{equation}
The angular integral enforces angular momentum conservation.
We introduce $n := n_i$ and $m := m_i$ and define
\begin{equation}
\begin{aligned}
    n_{\rm f} &=: n + \delta + 1\\
    m_{\rm f} &=: m + \ell + \delta
\end{aligned}
\end{equation}
such that angular momentum conservation is fulfilled for any choice of $n$, $m$, $\ell$, and $\delta$. Note that Refs. \cite{Gullans2017,Cao2022} only considered $\delta=0$ case. With this
\begin{equation}
    \mu^+(\ell, \delta, n, m) = i^{\ell}  \sqrt{\frac{n! (n + \delta)!}{m! (m + \delta + \ell)!}}  \int_0^{\infty} \mathrm{d}x \, 2 x e^{-x^2} x^{2(m - n) + \ell} L_{n}^{m - n}(x^2) L_{n + \delta}^{m - n + \ell}(x^2).
\end{equation}

We plot $\mu^+$ as a function of $\delta$ for $\ell = 0, {+}1, {+}2$ in Fig.~{\ref{fig:SelectionRules}} for $n = 37$ and $m=5000$ (a) and $m = 500$ (b).
For $m = 5000$ the expected radial position of the electron is $\langle r \rangle_{n = 37, m = 5000} \approx 1\ \mu \mathrm{m}$ hence somewhat in the radial middle of the sample that has an overall radius of $R \approx 2\ \mu \mathrm{m}$. 
For $m = 500$ we have  $\langle r \rangle_{n = 37, m = 500} \approx 300\ \mathrm{nm}$, which is close to the radius of the inner contact.
For $\ell = 0$ the orthogonality of the Laguerre polynomials enforces $\delta = 0$ such that we get the well known selection rules \cite{wendler_ultrafast_2015}
\begin{equation}
    \mu^+(\ell = 0, \delta, n, m) = \delta_{\delta, 0} \Rightarrow \mu^+ = \delta_{n_{\rm f} - 1, n_{\rm i}} \delta_{m_{\rm f}, m_{\rm i}}
\end{equation}
and for $\mu^-$
\begin{equation}
    \mu^-(\ell = 0, \delta, n, m) = \delta_{\delta ,0} \Rightarrow \mu^- = \delta_{n_{\rm f}, n_{\rm i} - 1} \delta_{m_{\rm f}, m_{\rm i}}.
\end{equation}
For $\ell \neq 0$ we don't get strict Kronecker-Delta-like selection rules but instead, multiple transitions are in principle allowed.
However, as can be seen in Fig.~\ref{fig:SelectionRules}, transitions with $\delta \neq 0$ are suppressed compared to the $\delta = 0$ transitions.
We illustrate the selection rules, only considering $\delta = 0$, in Fig.~\ref{fig:SelectionRules} (c).

In the main text, as a control experiment, we have inverted the $B$-field $\vec{B} \to - \vec{B}$.
It is therefore interesting to consider how the selection rules change in this case.
For deriving the selection rules as above, we have used the $B$-field to define the $z$-axis of the system.
Therefore the sign of the OAM of light as well as the handedness of the polarization is determined by whether they are aligned or anti-aligned with the magnetic field.
Inverting the $B$-field is therefore equivalent to inverting the OAM and the polarization.
Hence we can relate the transition matrix elements at magnetic field $\vec{B}$ and $-\vec{B}$ as
\begin{equation}
    \begin{aligned}
        \mu^+(\ell, \delta, n, m) \Big|_{\vec{B}} = \mu^-(-\ell, \delta, n, m)\Big|_{-\vec{B}}.
    \end{aligned}
\end{equation}
Based on this observation, when measuring the subtracted photo-current at magnetic field $\vec{B}$
\begin{equation}
    \Delta I\Big|_{\vec{B}} = I^{+\ell}\Big|_{\vec{B}} - I^{-\ell}\Big|_{\vec{B}}
\end{equation}
one expects this to change sign when inverting the magnetic field hence
\begin{equation}
    \Delta I\Big|_{\vec{B}} = -\Delta I\Big|_{-\vec{B}}.
\end{equation}
This is what we have used as a control experiment in the main text and indeed we observe the same sign reversal in the experiment.

%%%%%%%%%%%%%%%%%%%%%%%%%%%%%%%%%%%%%%%%%%%%%%%%%%%%%%%%%%%%%%%%%%%%%%%%%%%%%%%%%%%%%%%%%%%%%%%%%%%%%%%%%%%%%%%%%%%%%%%%%%%%%%%%%%%%%%%%%%%%%%%%%%%%%%%%%%%%%%
\newpage
\section{S8. Relation between expected radius and quantum numbers $n$ and $m$}
\begin{figure}[h]
    \centering
    \includegraphics{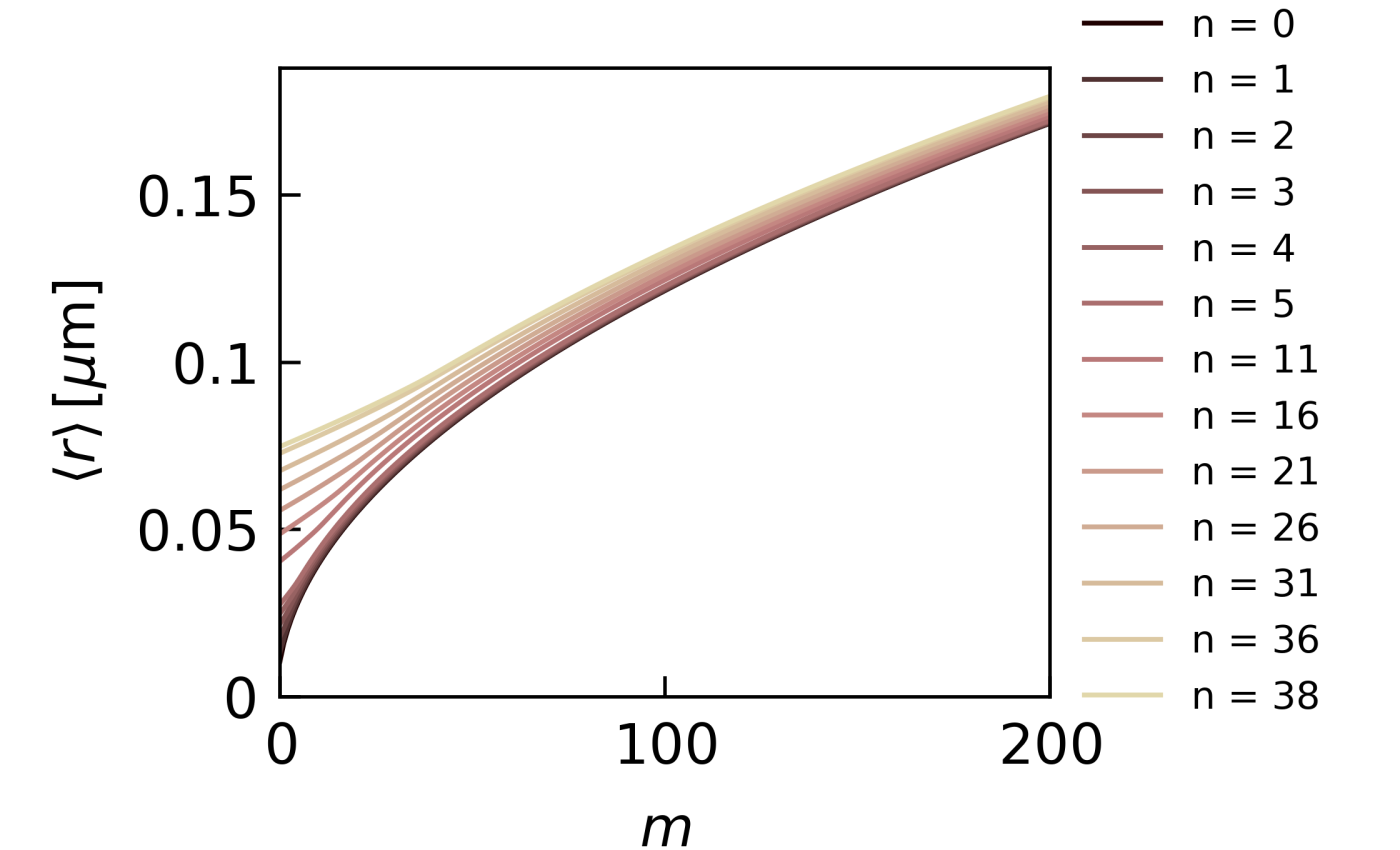}
    \caption{Expected radial position $\langle r \rangle$ according to Eq.~(\ref{eq:radius2Bands}) as a function of quantum number $m$ in different LLs $n$.}
    \label{fig:radiusDependence}
\end{figure}

In this part, we compute the expected radius of the electronic wavefunctions in graphene in the quantum Hall regime.
In spinor representation, they can be computed as
\begin{equation}
    \langle r \rangle = \begin{cases} \langle n=0, m| r | n=0, m\rangle \, ; \,\,\, n=0 \\ \frac{1}{2} \left( \langle n, m| r | n, m\rangle + \langle n-1, m| r | n-1, m\rangle  \right) \, ; \,\,\, n \neq 0 \end{cases}.
    \label{eq:radius2Bands}
\end{equation}
The wave functions $\phi_{n,m}(r) = \langle r, \phi | n, m \rangle$ read \cite{wendler_ultrafast_2015, goerbig_electronic_2011}  
\begin{equation}
  \phi_{n, m}(r, \varphi) = \frac{i^{|n - m|}}{\sqrt{2 \pi} \ell_{\rm B}} \sqrt{\frac{\text{min}(n, m)!}{\text{max}(n, m)!}} e^{-\frac{r^2}{4 \ell_{\rm B}^2}} e^{i \varphi |n - m|} r^{|n - m|} L_{\text{min}(n, m)}^{|n - m|} \left(\frac{r^2}{2 \ell_{\rm B}}\right)
\end{equation}
and are normalized according to 
\begin{equation}
    \int_{\mathbb R^2} \text{d}\vec{r} \, \phi_{n,m}(\vec{r}) \phi^*_{n',m'}(\vec{r}) = \delta_{n, n'} \delta_{m, m'}.
\end{equation}
We show the expected radii as a function of the quantum number $m$ for different values of $n$ in Fig.~\ref{fig:radiusDependence}.
For $n = 0$ the expected position approximately coincides with the position of the guiding center $r_{\mathrm{Guide}} = \ell_{\rm B} \sqrt{2 m + 1}$ while for larger $n$ one gets larger radii as expected.

%%%%%%%%%%%%%%%%%%%%%%%%%%%%%%%%%%%%%%%%%%%%%%%%%%%%%%%%%%%%%%%%%%%%%%%%%%%%%%%%%%%%%%%%%%%%%%%%%%%%%%%%%%%%%%%%%%%%%%%%%%%%%%%%%%%%%%%%%%%%%%%%%%%%%%%%%%%%%%
\newpage

\section{S9. Ideal interacting quantum Hall OAM response}
The above argument can be generalized to understand the OAM PC in the presence of Coulomb interactions. Let us consider photons incident 
on disorder-free Graphene in a perfect Corbino geometry with a magnetic field and Coulomb interactions. The Coulomb interaction is key to helping the carriers to thermalize and our discussion includes relaxation processes such as Auger. Let us further assume that the bulk is gapped and there are two edges, i.e., an inner and outer edge, which for now we will assume to be isolated from contacts. 

Given the circular symmetry, the angular momentum is the key quantum number to think about. In fact, in this ideal situation, energy $E$ and angular momentum $L$ are the only two conserved quantities. For a gapped bulk, the system thermalizes to a state with a low density of bulk quasiparticles. This means that the edge charge $Q$ is also a conserved quantity. As a consequence, the final state (after the absorption of some number of photons) is characterized by edges with quantum numbers $L$, $E$, and $Q$.

The edge theory is more conventionally described in terms of momentum $k$ rather than angular momentum. Note that in a magnetic field, momentum is shifted by the vector potential at radius $R$ of the edge is $k_{\rm{F}}=A=B R/2$. Transferring charge $Q$ to the edge changes momentum by $Q k_{\rm{F}}$. Adding energy through particle-hole excitations changes momentum by $E/v$ where $v$ is the edge velocity. The change in angular momentum is $L=R(Q k_{\rm{F}} +E/v)=R(Q B R/2+E/v)$. This is consistent with the wave-functions of LLs at angular momentum $m$ seen in the last section being at a radius $\sim\sqrt{2 m/B}$.

Photons supply both OAM angular momentum $l$ and energy $E_{\rm{ph}}$. Considering angular momentum and ignoring the contribution of the photon energy, the total charge transferred is $\delta Q_L=2 N_{\rm{ph}}l/B R^2= 2 N_{\rm{ph}}l (l_B/R)^2$. The photon energy contribution turns out to be independent of OAM and therefore even in magnetic field. The charge transfer expression matches that from the LL wave-function picture in the previous section and shows that interactions (at least in the absence of phonons and disorder) do not affect the OAM PC, which is a result of angular momentum pumping.

%%%%%%%%%%%%%%%%%%%%%%%%%%%%%%%%%%%%%%%%%%%%%%%%%%%%%%%%%%%%%%%%%%%%%%%%%%%%%%%%%%%%%%%%%%%%%%%%%%%%%%%%%%%%%%%%%%%%%%%%%%%%%%
\newpage

\section{S10. PC sign oscillation: LLs and injection current picture}

\textbf{\textit{LLs picture:}}
As described in a simplified picture earlier (Fig. 1 main text), in the quantum Hall regime, light with non-zero OAM can induce optical transitions between the LLs such that the spatial extent of these states depends on the associated angular quantum number $m$, where the radius of LLs increases with $m$. In our Corbino device, depending on the helicity of light, this spatial expansion (shrinking) of the carriers' wavefunction manifests itself as an outward or inward radial current.

This intuitive picture for the PC generation due to the radial modulation of the electron wavefunction caused by the optical vortex does not immediately suggest PC oscillations as a function of gate voltage ($V_{\rm{g}}$), which
are seen in Fig. 4 of the main text. To understand this effect, one must note that the
PC is actually the result of optical excitation and subsequent
carrier relaxation. As known from earlier studies \mbox{\cite{Cao2022,Nazin2010}} the position of 
$E_{\rm{F}}$ within a LL, controlled by $V_{\rm{g}}$,
gives rise to a relaxation bottleneck for either electrons or holes. Due
to the opposite charges of the different carriers (electrons and holes), this leads to a change of
PC polarity. That is, PC oscillations upon sweeping
through a LL. While in the rectangular geometry experiment of Ref.\cite{Cao2022} this bottleneck argument relied on equal chirality
of electrons and holes at the edge of the sample, in the present
scenario, the direction of transport is determined by the relative
helicity of twisted light and magnetic field. Naively, one might expect
that the OAM would lead to a relative shift of electron and hole
position, i.e., to opposite shifts for the two types of carriers.
However, Coulomb attraction ensures that the shift is experienced by the
center of mass of the electron-hole-pair (see also Ref. \cite{Tobias2022} for the case of excitons). Hence, electrons and
holes are moved in the same direction which explains the observed current
oscillations.

Unfortunately, formalizing this picture, i.e., via optical Bloch
equations as has been done in Ref. \cite{Cao2022}, is
extremely complicated. Not only would it be necessary to explicitly
account for LL and orbital degrees of freedom, but the picture
also suggests that properly accounting for the Coulomb attraction between
electrons and holes will be crucial to capture the oscillations.
However, an alternative description is able to provide a formalism which
is suited to explain the observed behavior.

\textbf{\textit{Injection current picture:}}
In our system, since the photon frequency is in the optical regime (pump wavelength $940\ \rm{nm}=1.32\ \rm{eV}$ and $E_{n=37}-E_{n=-36}=1.315\ \rm{eV}$), one expects the LL spacing ($E_{n=37}-E_{n=36}=9\ \rm{meV}$) to be much smaller than the lifetime broadening of excited electrons and holes. Therefore, we can consider optically excited electrons and holes as approximately free Dirac particles. Beyond the dipole approximation, the OAM of the pump photons locally imparts momentum to the electrons and holes in the tangential direction (in the polar coordinates of the Corbino disk). Moreover, the Lorentz force from the external out-of-plane magnetic field imparts radial components to the velocity of the electron and holes. As detailed in the following section, the radial current of both the electrons and holes in this scenario is therefore given by the following 
\begin{equation} \label{Ir}
I_{\text{radial}} = -\frac{Bq}{k^4}\cos(2\theta)(\tau_e + \tau_h)
\end{equation}
where $B$ is the external out-of-plane magnetic field, $q$ is the local OAM-induced momentum, $\theta$ is the angle between the electron and hole's momentum $k$, and $\tau_e$ and $\tau_h$ are the scattering times of the electrons and holes, respectively. From the Dirac equation $k=\omega/(2v_D)$. The total induced current is
\begin{equation} \label{It}
I_{\text{meas}}=\int{d\theta W(\theta) I_{\text{radial}}(\theta)}
\end{equation}
where $W(\theta)$ is the weight of the current contribution from the electrons and holes moving at the angle $\theta$. In an ideal case, $W$ should be $\theta$-independent, leading to a vanishing of the measured current in our system. However, as explained later in the following section, $W$ can be a function of both $\theta$ and the longitudinal conductivity, which is a function of the carrier density (and therefore gate voltage $V_{\rm{g}}$) and $B$. This can potentially explain the observed $V_{\rm{g}}$-dependent polarity change of the measured PC.

%%%%%%%%%%%%%%%%%%%%%%%%%%%%%%%%%%%%%%%%%%%%%%%%%%%%%%%%%%%%%%%%%%%%%%%%%%%%%%%%%%%%%%%%%%%%%%%%%%%%%%%%%%%%%%%%%%%%%%%%%%%%%%

\section{S11. Theory of OAM photoresponse in the strong scattering regime}\label{OAMscattering}
The OAM field from light with frequency $\omega$ can be written as $A(\vec{r})\propto e^{i l\theta}$, where $(r,\theta)$ are the radial and angular coordinates. This can then be expanded in terms of a tangential coordinate $x_t=r\theta$ as $A(\vec{r})\propto e^{i l x_t/r}=e^{i q x_t}$ where $q=l/r$. Locally, in the Corbino geometry, we can approximate the OAM as a plane wave with wave-vector $q$.

\begin{figure}
    \centering    
    \includegraphics[width=0.4\columnwidth]{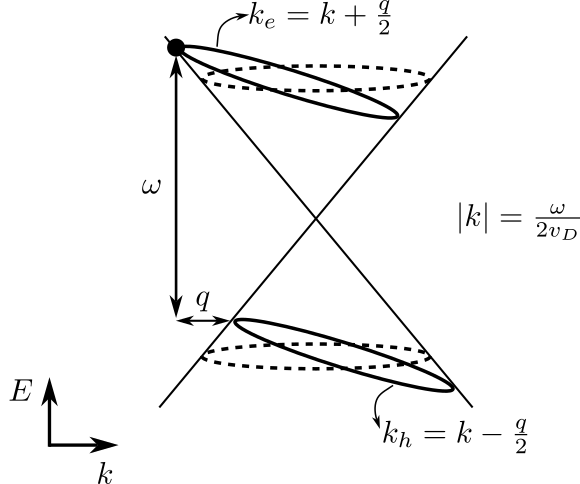}
    \caption{
    Optical excitation by finite wave-vector light in a band structure of Dirac electrons in graphene. 
    }
    \label{Diracbandstructure}
\end{figure}

Considering the absorption of the photon by Dirac electrons, at rate $R$ per unit area, shown in Fig.~\ref{Diracbandstructure}, leads to an electron-hole pair with momenta $k_e=k+q/2$ and $-k_h=-(k-q/2)$, respectively, where $|k|=\omega/2v_{\rm{D}}$ is independent of $q$. To check that this is correct note that the energy of such a pair is $|k+q/2|+|k-q/2|=2|k|+o(q^2)$.

\begin{figure}
    \centering    
    \includegraphics[width=0.4\columnwidth]{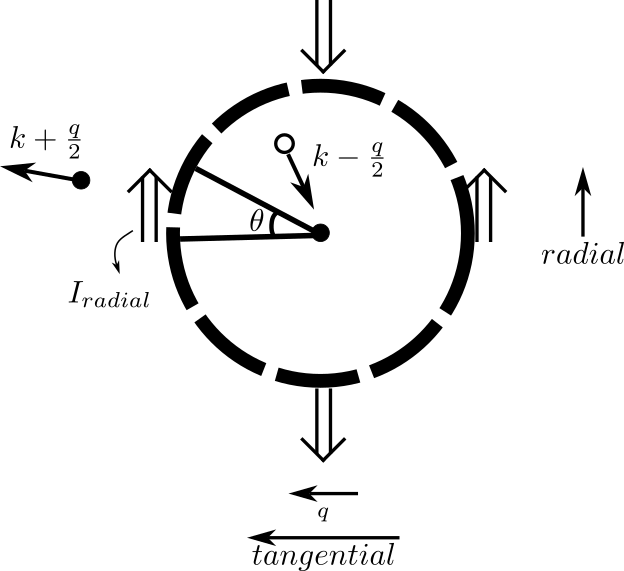}
    \caption{
    Wave-vector and velocity structure of electron-hole pairs in a magnetic field excited by light at wave-vector $q$, which is tangential in direction relative to the Corbino geometry. The dotted line shows the wave-vector $|k|=\omega/2 v_{\rm{D}}$ where the optical excitations originally occur. 
    }
    \label{Electronholestructure}
\end{figure}

Next, we consider (seen in Fig.~\ref{Electronholestructure}) the generated electron-hole pair in momentum space along with the velocity of each particle as well as the Lorentz-force induced momentum change. For a quasiparticle at momentum $k$ (note that we designate holes to be at the momentum of the missing electron), the velocity is $u_{\rm{e,h}}(k)=\pm v_{\rm{D}} \vec{k}/k$. The Lorentz-force induced momentum change is $B\times u_{\rm{e,h}}(k)\delta \tau_{\rm{e,h}}$ where $\tau_{\rm{e}}$ ($\tau_{\rm{h}}$) is the scattering time of electrons (holes). The average (over a scattering time) change in velocity induced by the Lorentz force turns out to be $\delta u_{\rm{e,h}}(k)=v_{\rm{D}}^2\tau_{\rm{e,h}}\frac{B\times\vec{k}}{2 k^2}$, which is the same for electrons and holes. The sign is independent of electrons or holes because both the velocity as well as the momentum change changes sign for both electrons or holes.

The Lorentz-force-induced current for electrons and holes at wave-vector $k$ would cancel (since the velocities are the same) except for the relative wave-vector $q$ induced by the OAM. This leads to an OAM-B-field induced velocity $\delta_2 u(k)=(q\cdot\nabla_k)\delta u(k)=\frac{q v_{\rm{D}}^2\tau B}{k^2}(\sin{2\theta},-\cos{2\theta})$ where $\theta$ is the angle of $\vec{k}$ relative to the tangential direction $\vec{q}$. This velocity is opposite for electrons and holes so that the electron and hole components of the current now add up. The relevant current in the Corbino geometry is the radial current $I_{\rm{rad}}=\frac{q_{\rm{e}}^2 R v_{\rm{D}}^2(\tau_{\rm{e}}^2+\tau_{\rm{h}}^2)q B}{k^2}\cos{2\theta}$ $=q_{\rm{e}} (R l_B^2)  (2\pi l)\{(\tau_{\rm{e}}^2+\tau_{\rm{h}}^2)\omega_c^2\}\cos{2\theta}$. The prefactor $2q_{\rm{e}}\pi l (R l_B^2)$ is the ideal expectation based on angular momentum. 

As shown in the figure, the radial component of the current flips sign with $\theta$, i.e., $I_{\rm{rad}}(\theta\sim 0,\pi)>0$ while $I_{\rm{rad}}(\theta\sim \pm \pi/2)<0$. In fact, the straightforward average over $\cos{2\theta}$ would vanish. However, for reasons we describe below, the total measured radial current $I_{\rm{meas}}=\int d\theta W(\theta)I_{\rm{rad}}(\theta)$ is a weighted average of the radial current where the presence of a $\cos{2\theta}$ Fourier component of $W(\theta)$ would lead to a non-vanishing measured current.

\begin{figure}
    \centering    
    \includegraphics[width=0.2\columnwidth]{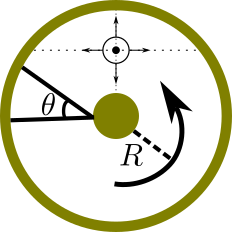}
    \caption{
    Scattering in Corbino geometry. Quasiparticles moving radially towards the edge, i.e., $\theta\sim \pm\pi/2$, have a shorter scattering time resulting in smaller B-field current though can more efficiently reach the edge because of ballistic transport relative to those moving tangentially, i.e., $\theta\sim 0,\pi$.
    }
    \label{ScatteringWeightfigure}
\end{figure}

To understand the origin of a non-trivial weight function $W(\theta)$, we note that $\theta$ represents the angle of the wave-vector $k$ (and hence velocity) relative to the tangential direction in the Corbino geometry at which the electron-hole pair is excited.  Electron-hole pairs at angles $\theta\sim \pm\pi/2$ are dominantly moving in the radial direction either towards or away from the Corbino edge. In contrast, pairs at $\theta\sim 0,\pi$ are moving tangentially. As clear from Fig.~\ref{ScatteringWeightfigure}, the formal set of electron/hole pairs and the associated current can hit the edge on a lower time scale. This reduces the scattering time $\tau_{e,h}$ for such quasiparticles. Thus, the weight function in the radial direction $W(\theta\sim \pm\pi/2)\sim (\tau_{\rm{e}}^{(r)}+\tau_{\rm{h}}^{(r)})/(\tau_{\rm{e}}+\tau_{\rm{h}})$, where $\tau^{(r)}_{\rm{e,h}}$ is the quasiparticle scattering time for particles moving in the radial direction and $\tau_{\rm{e,h}}$ is the scattering time of the quasi-particles averaged over the Fermi surface. In contrast, the tangentially moving quasiparticles have an averaged scattering time, but have a contribution to the edge conductance that is suppressed by the longitudinal conductivity $\sigma_{xx,W}$ of warm carriers. Note that the relevant conductivity is that of electron-hole pairs that are relaxing from the high optical energy scale to the ground state. The relevant weight function in the tangential direction, which arises from a current divider effect between the bulk and contact resistance, is $W(\theta\sim 0,\pi)\propto R_{\rm{contact}}\sigma_{xx,W}$, where $R_{\rm{contact}}$ is 
the contact resistance and $\sigma_{xx,W}$ is the two-dimensional conductivity (which has the same units as $R_{\rm{contact}}^{-1}$). This is a rough qualitative estimate that may be refined by more detailed modeling of the conductivity. However, we expect the key feature of dependence on bulk conductivity to be robust. 

The form of the weight function discussed above can predict a switching of sign with changing gate voltage of the measured B-dependent part of the OAM PC. As the gate voltage changes, the longitudinal conductance $\sigma_{xx,W}$ is expected to change as one passes through LLs. This leads to modulation of $W(\theta\sim \pm \pi/2)$, while one can expect $W(\theta\sim 0)\lesssim 1$ to have minor variations. At the same time $I_{\rm{rad}}(\theta\sim 0,\pi)\sim -I_{\rm{rad}}(\theta\sim \pi/2)$. Therefore, $I_{\rm{meas}}\sim I_{\rm{rad}}(\theta\sim 0,\pi)[W(\theta\sim 0,\pi)-W(\theta\sim \pm\pi/2)]$ changes sign when the bulk conductivity changes from low to high or vice-versa as the gate voltage sweeps through LLs.

\newpage
%%%%%%%%%%%%%%%%%%%%%%%%%%%%%%%%%%%%%%%%%%%%%%%%%%%%%%%%%%%%%%%%%%%%%%%%%%%%%%%%%%%%%%%%%%%%%%%%%%%%%%%%%%%%%%%%%%%%%%%%%%%%%%
\bibliography{Main}

%%%%%%%%%%%%%%%%%%%%%%%%%%%%%%%%%%%%%%%%%%%%%%%%%%%%%%%%%%%%%%%%%%%%%%%%%%%%%%%%%%%%%%%%%%%%%%%%%%%%%%%%%%%%%%%%%%%%%%%%%%%%%%
\end{document}